\documentclass[12pt,preprint]{emulateapj}
\slugcomment{Accepted for publication in ApJ}
\shortauthors{Chatzikos et al.}
\shorttitle{Chandra Observation of Abell 2065}

\begin{document}

\title{CHANDRA OBSERVATION OF ABELL 2065: AN UNEQUAL MASS MERGER?}
\author{Marios Chatzikos\altaffilmark{1},
Craig L. Sarazin\altaffilmark{1},
and
Joshua C. Kempner\altaffilmark{2,3}}
\altaffiltext{1}{Department of Astronomy, University of Virginia,
P. O. Box 3818, Charlottesville, VA 22903-0818, USA;
mnc3z@virginia.edu, sarazin@virginia.edu}
\altaffiltext{2}{Department of Physics and Astronomy, Bowdoin College, 8800
College Station, Brunswick, ME 04011, USA; jkempner@bowdoin.edu}
\altaffiltext{3}{Harvard-Smithsonian Center for Astrophysics, 60 Garden
Street, Cambridge, MA 02138, USA}

\begin{abstract}
We present an analysis of a 41~ks {\it Chandra} observation
of the merging cluster Abell~2065 with the ACIS-I detector.
Previous observations with {\it ROSAT} and {\it ASCA} provided evidence for an
ongoing merger, but also suggested that there were two surviving
cooling cores, which were associated with the two cD galaxies in
the center of the cluster.
The {\it Chandra} observation reveals only one X-ray surface brightness
peak, which is associated with the more luminous, southern cD galaxy.
The gas related with that peak is cool and
displaced slightly from the position of the cD.
The data suggest that this cool material has formed a cold front.
On the other hand, in the higher spatial resolution {\it Chandra}
image, the second feature to the north is not associated with the
northern cD; rather, it appears to be a trail of gas behind the main
cD.
We argue that only one of the two cooling cores has survived the merger,
although it is possible that the northern cD may not have possessed a
cool core prior to the merger.
We use the cool core survival to constrain the kinematics of the merger
and we find an upper limit of $\lesssim$~1900~km~s$^{-1}$ for the merger
relative velocity.
A surface brightness discontinuity is found at $\sim$~140~kpc
from the southern cD; the Mach number for this feature
is ${\cal M} = 1.66^{+0.24}_{-0.32}$, although its nature (shock
or cold front) is not clear from the data.
We argue that Abell~2065 is an example of an unequal mass merger.
The more massive southern cluster has driven a shock into the ICM of
the infalling northern cluster, which has disrupted the cool core of
the latter, if one existed originally.
We estimate that core crossing occurred a few
hundred Myr ago, probably for the first time.
\end{abstract}

\section{Introduction}
\label{sec:intro}

Mergers of galaxy clusters are the most energetic phenomena in the
Universe since the Big Bang.
Within the context of the hierarchical structure formation scenario,
cluster mergers are the mechanisms by which galaxy clusters form.
In a major merger, two subclusters collide at velocities of
$\sim 2000$ km s$^{-1}$, with a release of gravitational
potential energy of $\sim 10^{64}$ ergs.
The gaseous halos of the colliding clusters are subject to
hydrodynamical shocks that dissipate a major portion of the
released energy by heating the intracluster medium (ICM).
Cluster merger shocks may also generate turbulence and magnetic
fields and may accelerate relativistic particles.

The interaction of cooling cores, which are found in the
centers of relatively relaxed galaxy clusters, and merger
shocks has been the subject of recent research.
\citet{GLRB+02} conducted two-dimensional numerical simulations
to study the impact of cluster mergers on the survival of cooling
cores for the case of head-on collisions.
They found that for two mass ratios of 16:1 and 4:1, the crucial
parameter that determines the fate of the cooling core of the
primary component is the ram pressure experienced by the infalling
subcluster.
Disruption occurs when a significant mass of the infalling gas reaches
the primary's core.
The subcluster gas can heat up the cooling gas and increase the
cooling time by a factor of 10--40 by processes such as displacement
of the cooling core from the potential well, merger shocks and
turbulent gas motions.
Furthermore, the initial cooling time is important, as the cooling
core may reestablish itself, if the final cooling time is less than 
the Hubble time.
A time lag of $\sim$~1--2 Gyr was also found between core passage
and cooling flow disruption.

On the other hand, \citet{Heinz03} performed numerical simulations
to examine the effects of merger shocks on cooling cores.
They concluded that the outer parts of cool cores are being ram
pressure stripped, while gas motions, induced inside the core,
can result in transport motions in the direction of the incoming
shock wave and may eventually lead to the formation of a cold front.

A mechanism to account for cooling flow heating was
proposed by \citet{FSW04}.
In their model, cluster mergers are thought to produce a
significant amount of turbulence in the cluster ICM, which
in turn generates sound waves that contribute to the heating
of the cooling flow and suspend cooling to low temperatures.

In this report, we present results of the X-ray analysis of the
cluster Abell~2065.
This is a nearby ($z=0.072$) richness class 2 cluster \citep{ACO89}
and one of the 10 galaxy clusters that make up the Corona Borealis
supercluster.
Two cD galaxies dominate the cluster center in the optical;
their radial velocities differ by $\sim$~600~km~s$^{-1}$
\citep{PGH}.

Previous X-ray studies with {\it ASCA} and {\it ROSAT} led to
marginal detections of two surface brightness peaks coincident
with the two cD galaxies observed at the cluster center
\citep{MSV99}.
The data indicated that the cluster has undergone a merger.
\citet{MSV99} suggested that the cooling cores of
the two cD galaxies may have survived the merger,
and they used this to put constraints on the gravitational
potential peaks of the two clusters.

The present work is focused on the study of merger signatures
that can shed light on the merger dynamics, as well as on the
nature of the central cool gas and its association to the
two central cDs.

The organization of the paper is as follows.
In \S~\ref{sec:data} we describe the observations and data
reduction procedures.
In \S~\ref{sec:images} we present X-ray images of the cluster
emission and compare them to optical and radio observations.
The X-ray spectral analysis and the derived temperature maps of the
cluster are presented in \S~\ref{sec:spec}.
The radial X-ray surface brightness profile is presented
in \S~\ref{sec:sbprof} and the spectral deprojection is
discussed in \S~\ref{sec:deproj}.
In \S~\ref{sec:disc}, a possible dynamical interpretation
of the merger is proposed.
We summarize our conclusions in \S~\ref{sec:sum}.

Throughout the paper, we use $H_0 = 72$ km~s$^{-1}$~Mpc$^{-1}$,
$\Omega_{\Lambda} = 0.73$ and $\Omega_M = 0.27$,
which corresponds to a linear scale of 1.336~kpc~arcsec$^{-1}$
at the redshift of the cluster.
When errors are reported, they represent 90\% confidence intervals.
Position angles (PAs) are measured from north to east.

\section{Observation and Data Reduction} \label{sec:data}

Abell 2065 was observed by {\it Chandra} with the ACIS-I detector on
2002 August 18 for 28~{\rm ks} (Obs.\ 1).
The observation was interrupted by a Solar flare and resumed on 2002
November 24 for another 22 ks (Obs. 2).
The telemetry of the data was performed in Very Faint mode.

Due to the considerable pointing and orientation offsets, the initial
reduction of the two datasets has been performed independently.
The data have been calibrated using the Calibration Database, 
CALDB\footnote{http://cxc.harvard.edu/caldb/} version 3.0.
The Chandra Interactive Analysis of Observations software package, 
CIAO\footnote{http://cxc.harvard.edu/ciao/} version 3.2, 
was used for the reduction and the analysis of the datasets.

The standard filtering has been performed, accepting only events with
{\it ASCA} grades 0,2,3,4 and 6, and excluding known bad pixels, chips
nodes and chip boundaries.
In order to reduce the background, the extended grades based on the
Very Faint data were used to filter out additional events which were
due mainly to particle background.
The data have been corrected for the charge-transfer inefficiency effect 
(CTI).
Spatial and temporal corrections for the QE degradation effect
were applied automatically by the software.

The ACIS-S2 chip was used to monitor the background, as it includes
less of the cluster emission than the ACIS-I chips.
The onset of the flare resulted in a high background rate in the
August observation (Obs.\ 1).
Time intervals that exhibited background count rate fluctuations greater than
1-$\sigma$ from the mean have been excluded, and this resulted in 8 ks
being discarded.
Even with this clipping of high background data, the mean count rate
during the remaining good time is about 30\% higher than the expected
quiescent value for this
period\footnote{http://cxc.harvard.edu/contrib/maxim/acisbg/data/README}. 
However, if the filtering criteria were tightened, nearly all of the
time in Obs.~1 was rejected.
Since we are mainly concerned with very bright regions, where very few
of the counts are due to background, we retained the resulting 20 ks of
Obs.~1, and adjusted the blank sky background files' observing length
(EXPOSURE keyword in the FITS file headers) for this observation to be
consistent with the higher background rate.

The background of Obs.\ 2 was much lower.
Following the same cleaning procedure as with Obs.\ 1, less than
2~{\rm ks} were discarded.
The resultant background rate is $\sim$~1\% higher 
than the quiescent mean for the period of the
observation\footnotemark[5].

In order to check the registration of the two observations,
point sources have been detected in each observation using
the CIAO WAVDETECT\footnotemark[2] algorithm.
The positions of the sources were correlated, and the photon positions
in the second observation were corrected for the mean offset.
The events lists of the two observations were then merged.
The merged events list was used for all of the analysis, except for
spectral fitting.
Spectra were extracted separately for the two observations, and were fit
by a common model.

To check the absolute astrometry of the X-ray data, WAVDETECT was used
to detect X-ray point sources in the merged events list.
The X-ray positions of these sources were compared to the positions of
optical/IR sources in the {\rm 2MASS} and {\rm USNO-B1.0} catalogues.
A very small shift to the astrometry of the X-ray image was made
($-0\farcs11, -0\farcs06$) in RA and DEC, respectively.

\section{X-ray Images} \label{sec:images}

\begin{figure*}
\plotone{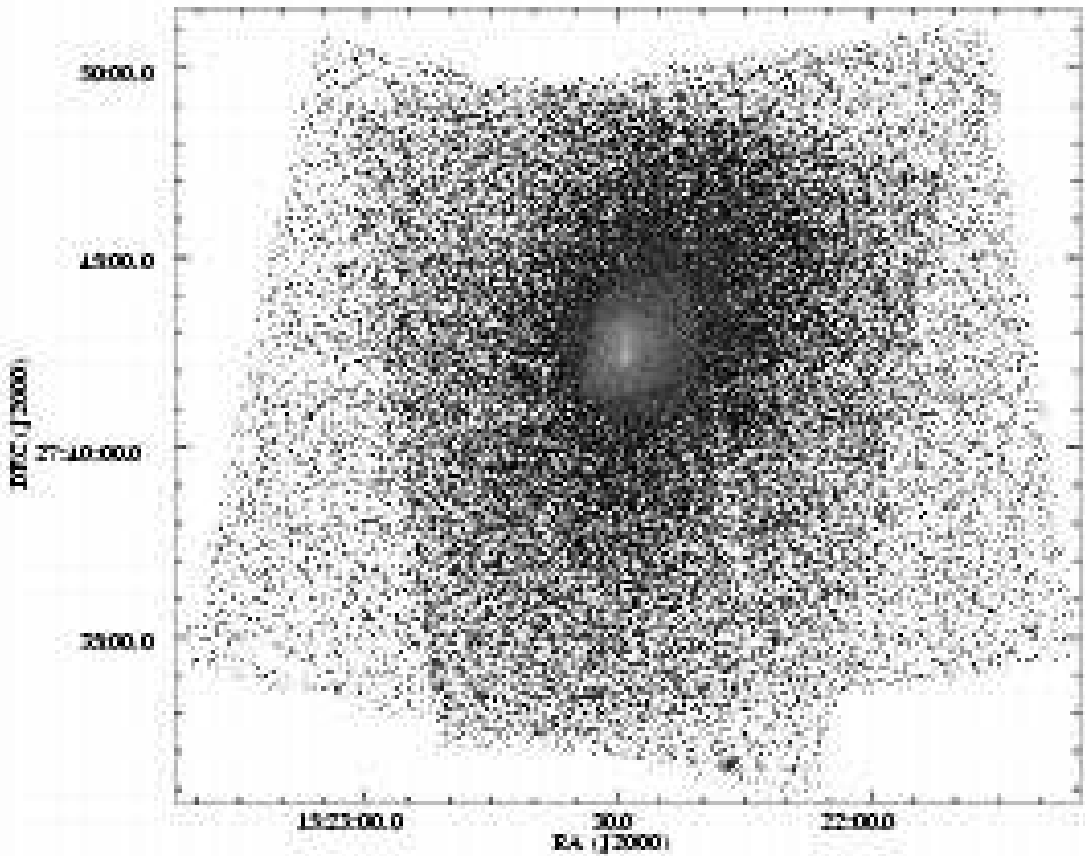}
\caption{Raw merged ACIS-I X-ray image of two observations in the 0.3--10
         keV band, uncorrected for background or exposure.
         The image was binned by a factor of 5 in RA and DEC
         and it is displayed in logarithmic scale.
         The irregular outline is the result of the different fields-of-view
         of the two observations.
         The two cross-like faint regions are the chip gaps between
         the four CCDs which make up the ACIS-I detector in each of
         the two observations.
         \label{fig:raw}}
\end{figure*}

\begin{figure*}
\plotone{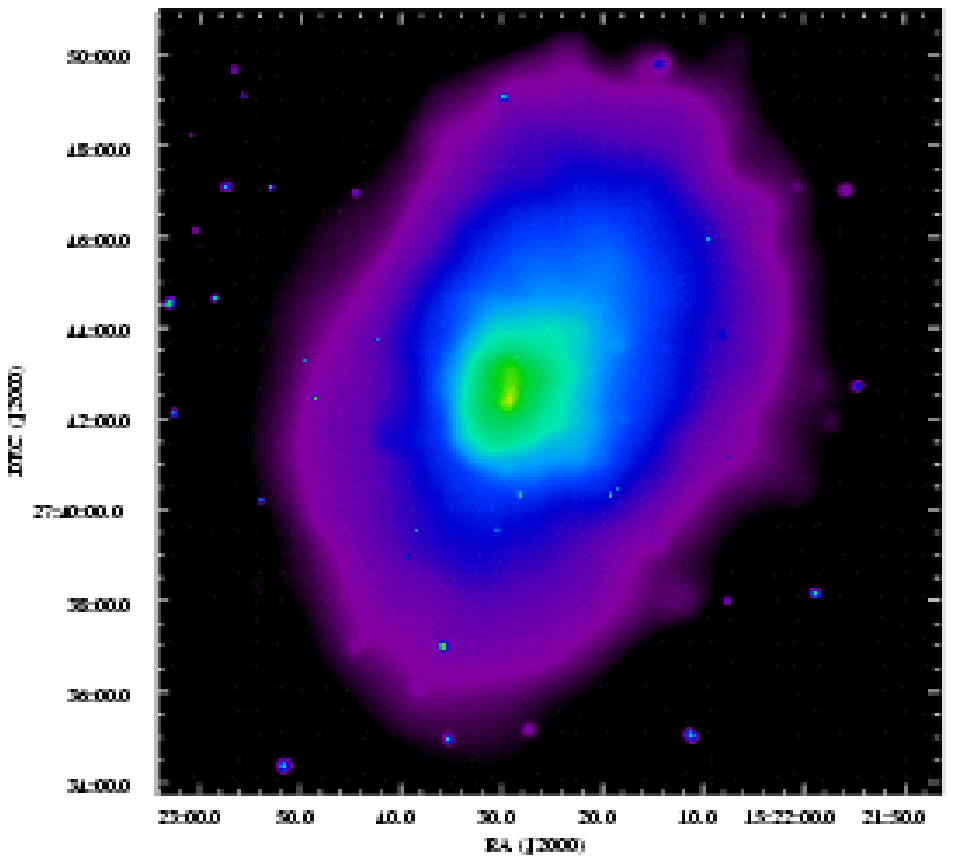}
\caption{Adaptively smoothed {\it Chandra} ACIS-I image of Abell 2065 in
         the energy band 0.3--10 keV, corrected for background and
         exposure.
         The image is displayed in logarithmic scale to emphasize low
         surface brightness emission.
         The faintest non-black regions have a surface brightness
         of $1.5 \times 10^{-6}$ counts~sec$^{-1}$~arcsec$^{-2}$.
         \label{fig:smooth}}
\end{figure*}

The raw X-ray image from the cleaned and merged observations is shown in
Figure~\ref{fig:raw}, uncorrected for background or exposure.
The image has been binned so that each pixel occupies 
2\farcs46$\times$2\farcs46 on the sky.
This image shows the complex field of view of the combined observations,
and the regions where chip gaps complicate the analysis.
An adaptively smoothed X-ray image in the 0.3--10 keV band, corrected for
background and exposure variations across the field of view,
is presented in Figure~\ref{fig:smooth}.
The image was smoothed to a minimum signal-to-noise per smoothing beam of
3-$\sigma$.

Overall, the cluster emission is elongated from the northwest to the
southeast.
At the center, the image shows a compact, bright region.
There is a more diffuse tail extending to the north at the center of the
image.
Although variations in the exposure confuse the region SE of the
cluster center, there is evidence for a bow-shaped surface brightness
discontinuity in this direction.
There is also an extension in the emission further to the SE.

\begin{figure*}
\epsscale{1.15}
\plottwo{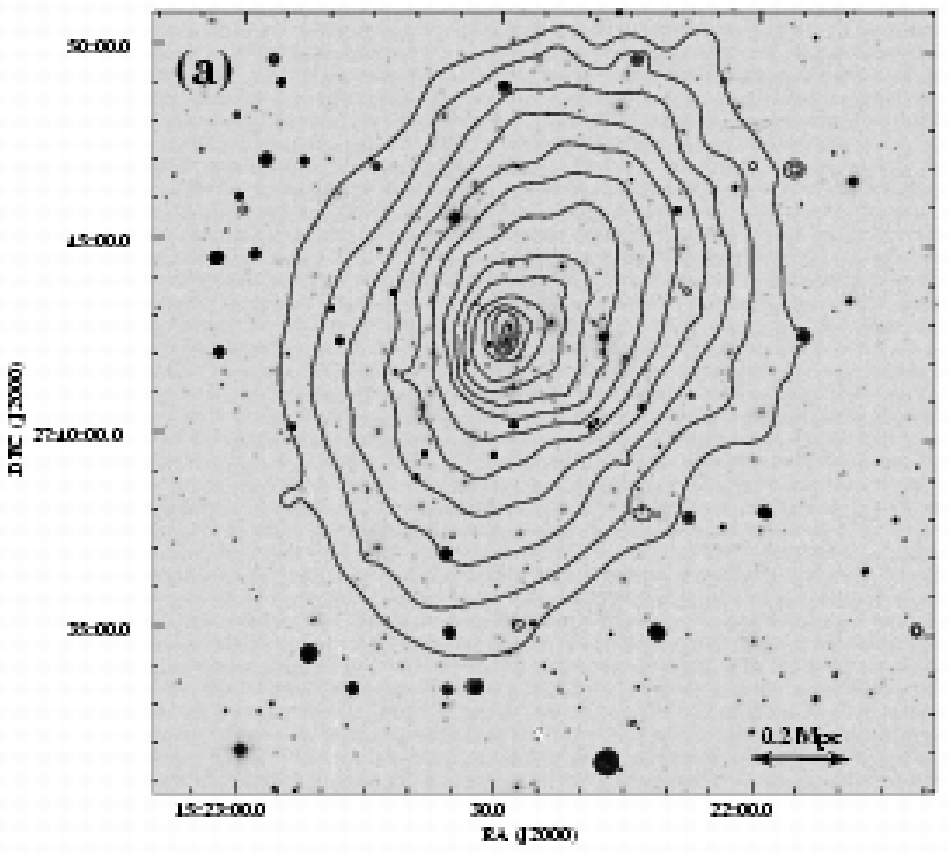}{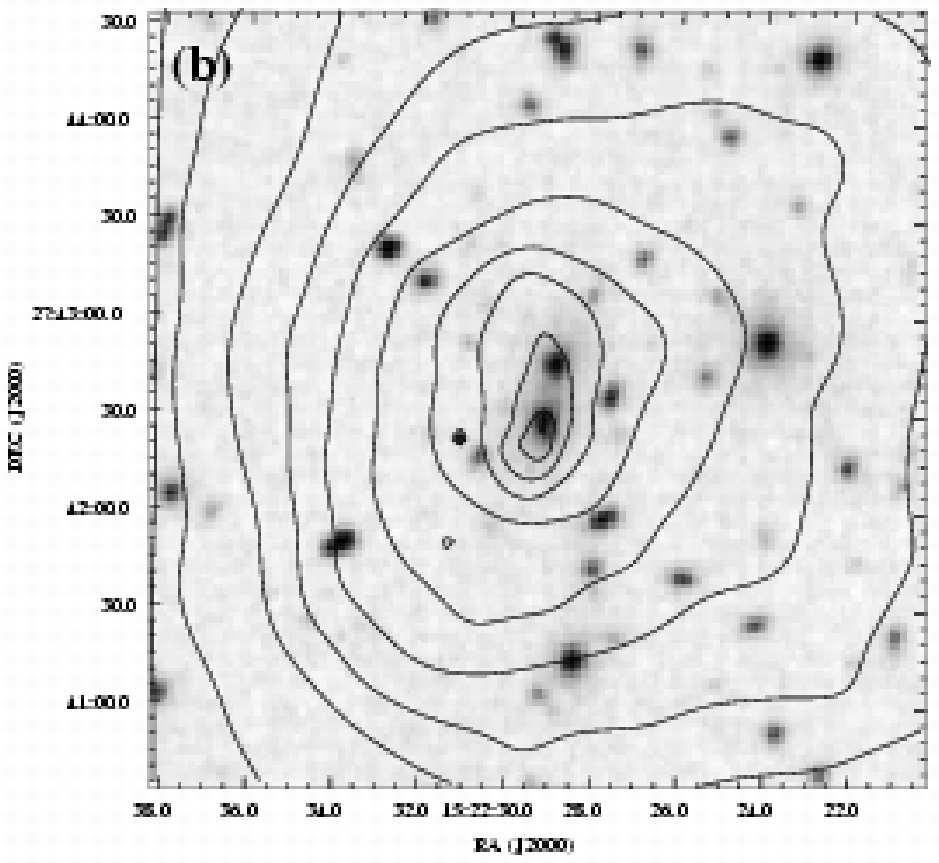}
\caption{{\it Chandra} X-ray contours overlaid on the Digitized Sky Survey
        image of Abell 2065.
        ({\it a}) 20\farcm5$\times$20\farcm5 field-of-view.
        The contours are spaced by a factor of $\sqrt{2}$ and they range
        from $1.5 \times 10^{-6}$ to $1.4 \times 10^{-4}$ 
        counts~sec$^{-1}$~arcsec$^{-2}$ for the diffuse gas.
        The brightest point sources reach surface brightness
        levels of $8.2 \times 10^{-3}$ counts~sec$^{-1}$~arcsec$^{-2}$.
        ({\it b}) The central 4\arcmin$\times$4\arcmin\ cluster region.
        Note that only the southern cD seems to be associated with 
        the X-ray surface brightness peak.
	\label{fig:dss}}
\end{figure*}

The optical DSS image of the cluster is presented in 
Figure~\ref{fig:dss}{\it a} overlaid with the X-ray 
contours from the adaptively smoothed X-ray image
(Fig.~\ref{fig:smooth}).
The central cluster region is shown in Fig.~\ref{fig:dss}{\it b}.
The X-ray peak at the cluster center is displaced by 2\farcs4 
from the position of the southern cD galaxy in the cluster.

\begin{figure*}
\epsscale{0.75}
\plotone{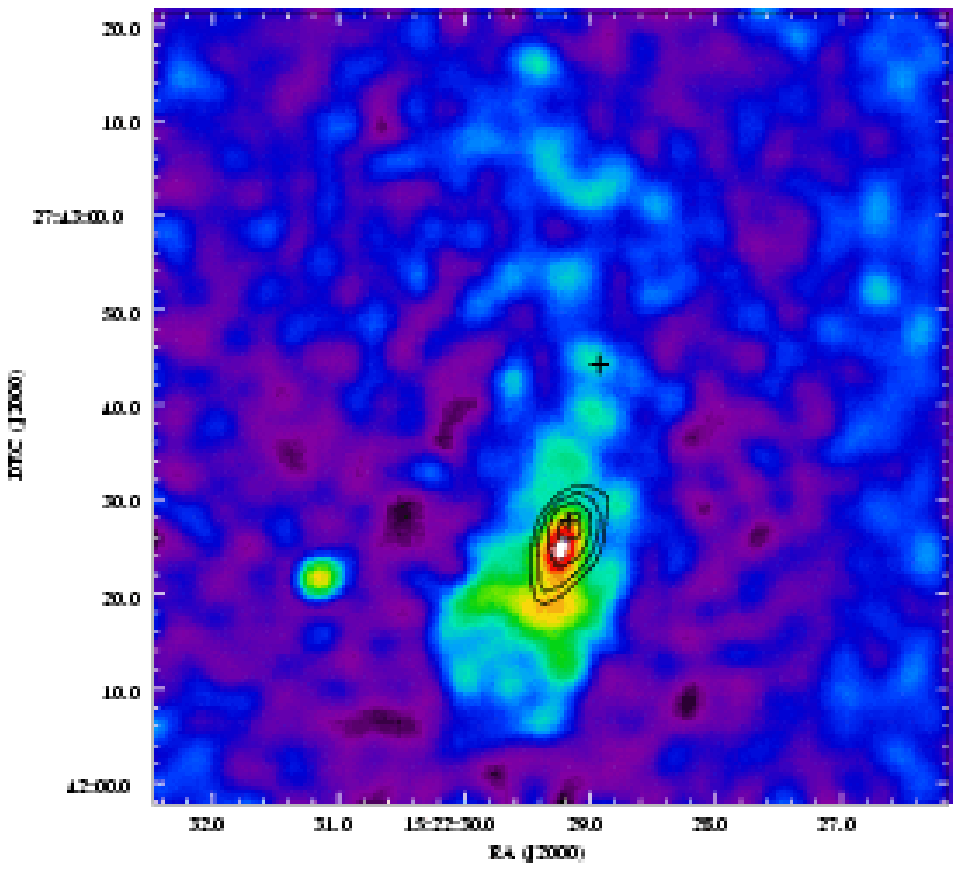}
\caption{Unsharp masked image of the central
         $\sim$~84\arcsec$\times$84\arcsec\ region of Abell 2065.
         The image was created by smoothing the 0.3--10~keV total
         (including the background) count rate image with two gaussian
         kernels, $\sigma=$~1\farcs476 and $\sigma=$~14\farcs76,
         and subtracting the first smoothed image from the second.
         The crosses indicate the positions of the centers of the two
         cD galaxies (2MASX~J15222892+2742441 and 2MASS~J15222917+2742275).
         The contours show the FIRST 1.4~GHz radio surface brightness.
         The lowest contour corresponds to a flux of 0.2 mJy per beam,
         and the contours are spaced by a factor of $\sqrt{2}$.
         The bright source to the E of the cluster center has no optical
         counterpart.
         \label{fig:radio}}
\end{figure*}

Following \citet{2003MNRAS.344L..43F}, we applied an unsharp
masking technique to reveal the structure in our data. 
Figure~\ref{fig:radio} shows an image of the central
$\sim 84\arcsec \times 84\arcsec$ region of Abell~2065
produced by this technique.
A 0.3--10~keV exposure-corrected image was smoothed with two
gaussian kernels of 1\farcs476 and 14\farcs76 (3 and 30 pixels
on the ACIS-I detector), respectively;
Fig.~\ref{fig:radio} displays the difference between the two
smoothed images.
This particular set of kernels was selected from a range of
sets, as they yield the highest degree of structure in the
image; smaller kernels tend to unresolve the largest scale
structure, while larger ones tend to wash out the high-resolution
features of the image.
Accurate positions of the centers of the two cDs have been
determined from the 2MASS catalogue and they are marked with
crosses in Figure~\ref{fig:radio}.
The position of the southern cD is close to the brightest
region of X-ray emission in the cluster.
The brightest X-rays come from a point just south of the center
of the southern cD, and there is an elongated extension of the
X-ray emission running about 25\arcsec\ further to the south.
There is also a more diffuse tail of emission to the north of the
southern cD extending about 45\arcsec.
The northern cD galaxy is located along the western edge of this
tail, but does not correspond to a surface brightness peak or any
other interesting feature in the tail.

Based on much lower resolution {\it ROSAT} images, \citet{MSV99} noted
that the X-ray emission at the center of Abell 2065 was marginally
resolved into two peaks;
again, at the lower resolution of the {\it ROSAT} data, these were
consistent with being located at the positions of the two cD galaxies.
\citet{MSV99} argued that the two peaks were two cool cores associated
with separate subclusters with the two cDs at their centers, and
that these two subclusters had merged.
They suggested that the cool cores were still bound to the two cD
galaxies, and drew conclusions about the central concentrations of the
dark matter potentials of the two cDs.
With our higher resolution {\it Chandra} data, the southern peak is indeed
associated with the southern cD, but the northern peak appears to
correspond to the more diffuse curved tail.
This is not centered on the northern cD galaxy.
Thus, only one of the cDs has retained a bound cool core,
although the northern cD may not have had a cool core prior
to the merger.

The morphology of the northern tail and its apparent connection with
the southern cD suggests that the tail may be gas which has been
stripped from this cD.
Alternatively, it is possible that it was removed from the northern
cD, which now retains very little hot gas.

The southern cD galaxy (2MASS J15222917+2742275) is also a faint
radio source (3.09~mJy at 1.4~GHz).
In Figure~\ref{fig:radio}, the radio contours from the FIRST survey are
superposed on the unsharp-masked X-ray image of the center of the
cluster.
The radio source is small and the core is
not very well resolved in the FIRST data,
although the extent to the south is much
larger than the FIRST beam size.
The peak in the radio surface brightness is nearly coincident with the
center of the cD galaxy and the peak in the X-ray surface brightness.
The radio source is extended to the SE of the center of the cD
galaxy.
The radio extension corresponds closely to a similar extension in the
X-ray emission.
If both are due to ram pressure as the southern cD moves relative to
the surrounding intracluster gas, this would indicate that the galaxy
is now moving to the NW.
If this is the case, then the more extended tail to the north is
due to gas removed from the northern cD.
On the other hand, the radio/X-ray extension to the south might be due
to buoyancy, or to more complex gas motions in the core of the cluster.

\section{Spectral Analysis}\label{sec:spec}

We extracted spectra of selected regions in the {\it Chandra} image
separately for the two datasets.
Flux weighted ancillary files and response matrix files were calculated
with the CIAO tools MKWARF and MKRMF.
Because in CIAO~3.2 the tool MKRMF did not work properly with the
latest calibration files (CALDB 3.0), the data were reduced anew
using the previous calibration (CALDB 2.8); the blank sky background
files were also processed to match the observations.
The calibration included the temporal and spatial variation of the
soft X-ray quantum efficiency degradation due to the build-up of
material on the optical blocking filter.
The data were screened using the same good time intervals
of low background brightness as in \S~\ref{sec:data}.
The spectra for each dataset were independently grouped, so that each bin
contained at least 25 counts.
In both observations, the spectrum at energies below 0.7~keV
is affected by sharp effective area and quantum efficiency
variations, while at energies above 6.5~keV the background
is dominant; thus we restrict our spectral extraction in this
photon energy range 0.7--6.5~keV.
The spectra from the two datasets were simultaneously
fit in this range using the same spectral model.
In most cases, the spectra were fit with the WABS model for the
absorption and the APEC model \citep{apec} for the thermal emission.
In most fits, the absorbing column density was held fixed to the
Galactic value of $N_H = 2.9 \times 10^{20}$ cm$^{-2}$.
The fitting was performed with {\rm XSPEC} \citep{XSPEC}, version 11.3.

\subsection{Global X-ray Spectrum} \label{sec:spec_global}

First, the spectrum was extracted for the largest elliptical region which
roughly followed the shape of the X-ray isophotes and which fell
completely within the fields-of-view of both observations;
this was an elliptical region centered on the southern cD galaxy, with
semimajor and semiminor axes of 6\farcm3 and 4\farcm9, respectively,
aligned with the orientation of the cluster emission (PA of the major
axis at $-20\degr$). 
The background was determined from the blank sky background files
discussed in \S~\ref{sec:data}.

\begin{figure}
\epsscale{0.8}
\rotatebox{-90}{\plotone{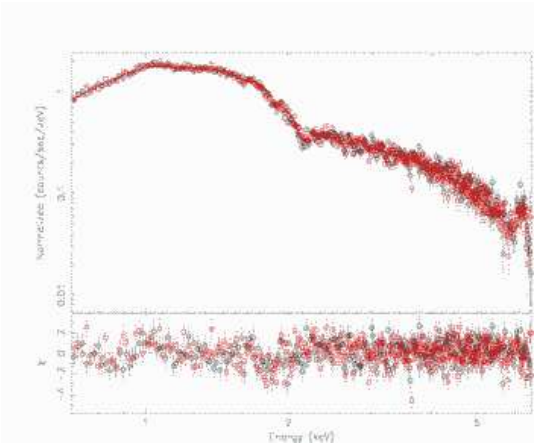}}
\caption{Global X-ray spectrum of Abell 2065, with the best-fit single
         temperature APEC model attenuated by Galactic absorption.
         Circles (black) and squares (red) represent Obs.~1 and Obs.~2
         spectra, respectively. {\it [See electronic edition of the Journal
         for a color version of this figure.]}
         \label{fig:spec_global}}
\end{figure}

Table \ref{tab:spec_global} summarizes the resulting spectral fits.
A single temperature model with Galactic absorption provided a good fit
to the spectrum, with a value of $\chi^2$ per degree-of-freedom, $\chi^2_r$,
very close to unity.
The observed spectrum and best-fit model are shown in
Figure~\ref{fig:spec_global}.
We tested the effect of allowing the absorption column density to vary;
this marginally improved the fit, and yielded a column density consistent
within the errors with the Galactic value (here and elsewhere, the F-test
has been used to compare the fits).
Therefore, in the following, we keep the
column density fixed at the Galactic value.
The best fit model shows a residual structure at $\sim$~2 keV;
it is known that this is generally due to the rapid variation in
{\it Chandra}'s effective area near the Ir edge of the telescope's mirror.
We refit the spectrum removing this region (1.8--2.3 keV).
This improved the fit, but the model parameters were
essentially identical to those including the Ir edge
region.

In \S~\ref{sec:spec_tmap} below, we will see that the cluster contains
gas with an extremely wide range of temperatures.
Thus, it is somewhat surprising that the global cluster spectrum is well
fit by a single temperature model, and so we tried a two temperature model.
The fit was improved, but the second thermal component was essentially
unconstrained, having a lower limit of about 10~keV.
This indicates that the cluster contains some very hot gas, but the global
spectrum is not very useful for characterizing this gas.

Previous {\it ROSAT} and {\it ASCA} images and spectra suggested that
Abell 2065 contained a weak cooling flow \citep{MSV99, MFSV+98, Peres, Whi00}.
Thus, we also tried a model of a single temperature cluster thermal
emission combined with a cooling flow component whose upper temperature
was fixed at the cluster ambient temperature and whose lower temperature
was 0.08 keV.
The abundance in the cooling flow component was fixed to that of the ambient
cluster gas.
XSPEC's MKCFLOW, which was used to model the cool core, utilizes the
MEKAL atomic library \citep{mekal} to calculate atomic line emission.
In order to use consistent atomic data between the ambient thermal
emission and the cool core model components, the MEKAL model was
employed to model the cluster component.
We first tested what effect the use of the MEKAL model had on the best-fit
parameter values as determined with the APEC model.
The fit was not significantly improved, and the cluster temperature and
abundance were consistent within the errors with the values obtained with APEC. 

Combining the thermal and cooling flow components did not produce
a better fit to the spectrum, and the best-fit cluster temperature
and abundance values were consistent with those derived from the
MEKAL model alone.
The cooling rate for the model was
$\sim$~4$^{+11}_{-4}$~M$_{\sun}$~yr$^{-1}$, which is in agreement with
the measurement of \citet{Peres}, but also consistent with no cooling.

The best-fit single temperature model is consistent with results from
the {\it ASCA} spectrum.
\citet{MFSV+98} found a best-fit single temperature of $kT = 5.4 \pm0.3$
keV; almost exactly the same result was derived by \citet{Whi00} and
\citet{IPB+02}.
On the other hand, \citet{DSJ+93} derived a temperature of
$kT = 8.4^{+3.3}_{-1.8}$ keV from the {\it Einstein} MPC detector.
Both our temperature map (\S~\ref{sec:spec_tmap}, below) and the previous
temperature map from {\it ASCA} \citep{MSV99} show that the outer, southern
parts of the cluster are very hot.
Thus, the higher {\it Einstein} MPC average temperature probably
results from the very large field of view and a very hard X-ray response
of this instrument.

\begin{deluxetable*}{lcccccccc}
\tabletypesize{\rm\small}
\tablecaption{Models Fit to the Global X-ray Spectrum
              \label{tab:spec_global}}
\tablehead
  {
  &  &  &
 \multicolumn{2}{c}{1st Component}  &  &
 \multicolumn{3}{c}{2nd Component}  \\
 \cline{4-5} \cline{7-9}\\
   {\scriptsize Model}       & 
   $\chi^2_r$  & 
   $N_H$       &
   $kT$        & 
   Abundance   &
               &
   $kT$        & 
   Abundance   &
   $\dot{M}$  \\
   {\scriptsize (WABS *)}&
                        &
   ($10^{20}$ cm$^{-2}$)&
   (keV)                &
   (Solar Units)        &
                        &
   (keV)                &
   (Solar Units)        &
   $M_{\sun}$ yr$^{-1}$
  }
\startdata
{\rm\scriptsize APEC}
                  &  852/731=1.166
                  &  $2.9$ 
                  &  $5.520\pm0.143$
                  &  $0.309\pm0.039$
                  &  
                  &  \nodata 
                  &  \nodata
                  &  \nodata \\

{\rm\scriptsize APEC}
                  &  851/730=1.165
                  &  $2.366\pm 0.676$
                  &  $5.627^{+0.200}_{-0.198}$
                  &  $0.310^{+0.040}_{-0.039}$
                  &
                  &  \nodata
                  &  \nodata
                  &  \nodata \\

{\rm\scriptsize APEC --- Ir region}
                  &  725/661=1.097
                  &  $2.9$
                  &  $5.513^{+0.143}_{-0.141}$
                  &  $0.310\pm 0.039$
                  &
                  &  \nodata
                  &  \nodata
                  &  \nodata \\

{\rm\scriptsize APEC+APEC} 
                  &  817/728=1.122
                  &  $2.9$
                  &  $4.035^{+0.217}_{-0.288}$
                  &  $0.459^{+0.063}$
                  &
                  &  $11.547_{-1.869}$
                  &  $0$
                  &  \nodata \\

{\rm\scriptsize MEKAL}
                  &  839/731=1.148
                  &  $2.9$
                  &  $5.541^{+0.125}_{-0.120}$
                  &  $0.282\pm0.034$
                  &
                  &  \nodata
                  &  \nodata
                  &  \nodata \\

{\rm\scriptsize MEKAL+MKCFLOW}
                  &  839/730=1.149
                  &  $2.9$
                  &  $5.609^{+0.238}_{-0.179}$
                  &  $0.282\pm0.034$
                  &
                  &  \nodata 
                  &  $0.282$
                  &  $4^{+11}_{-4}$
\enddata
\end{deluxetable*}

\subsection{Temperature Maps\label{sec:spec_tmap}}

\begin{figure*}
\epsscale{1.15}
\plottwo{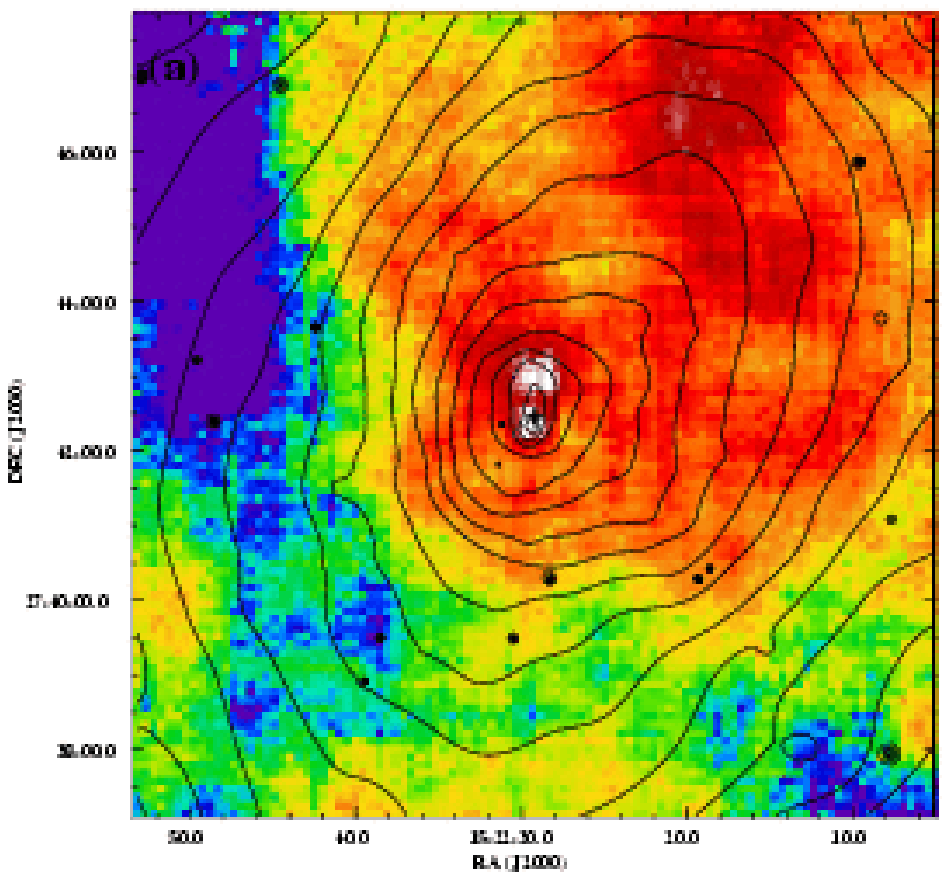}{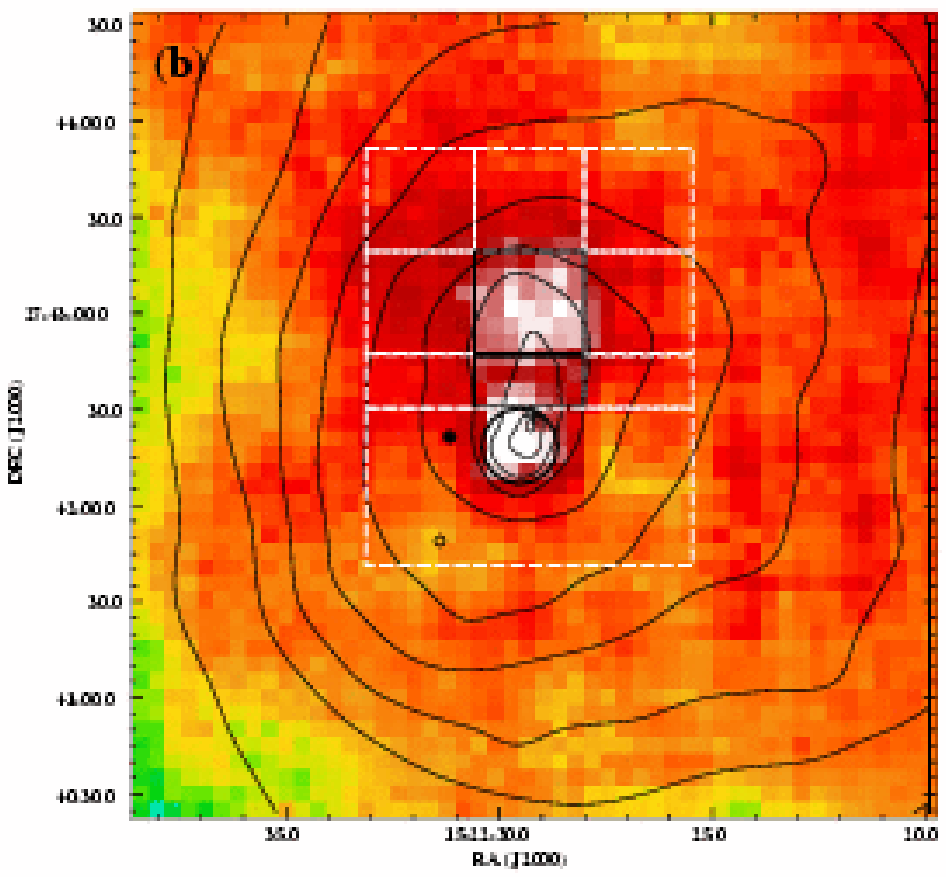}
\plottwo{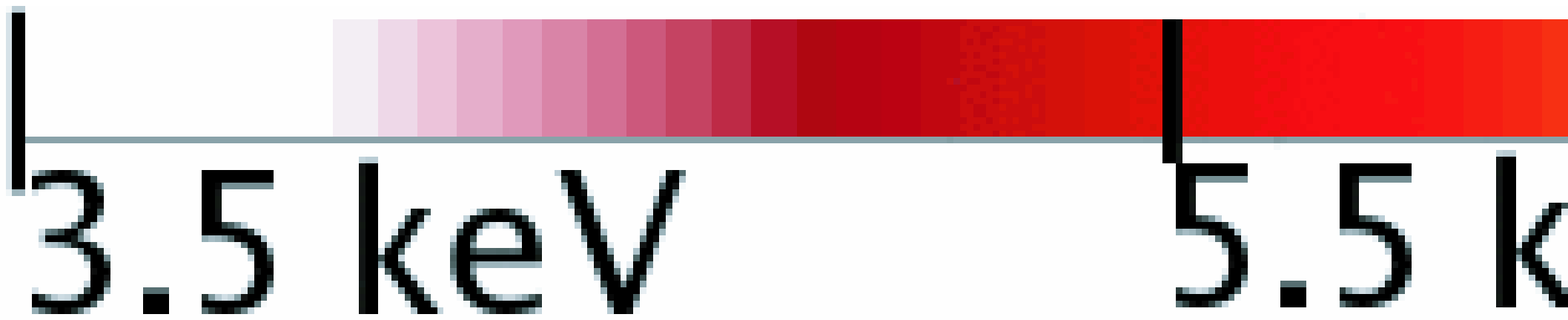}{f6e.eps}
\plottwo{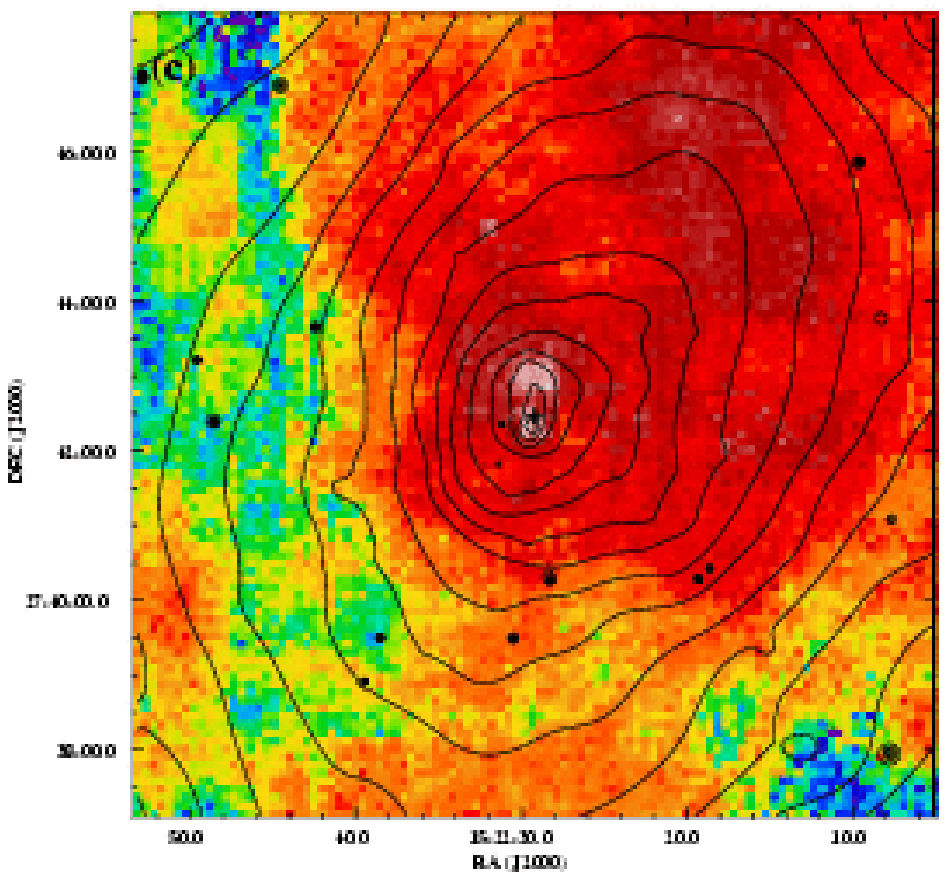}{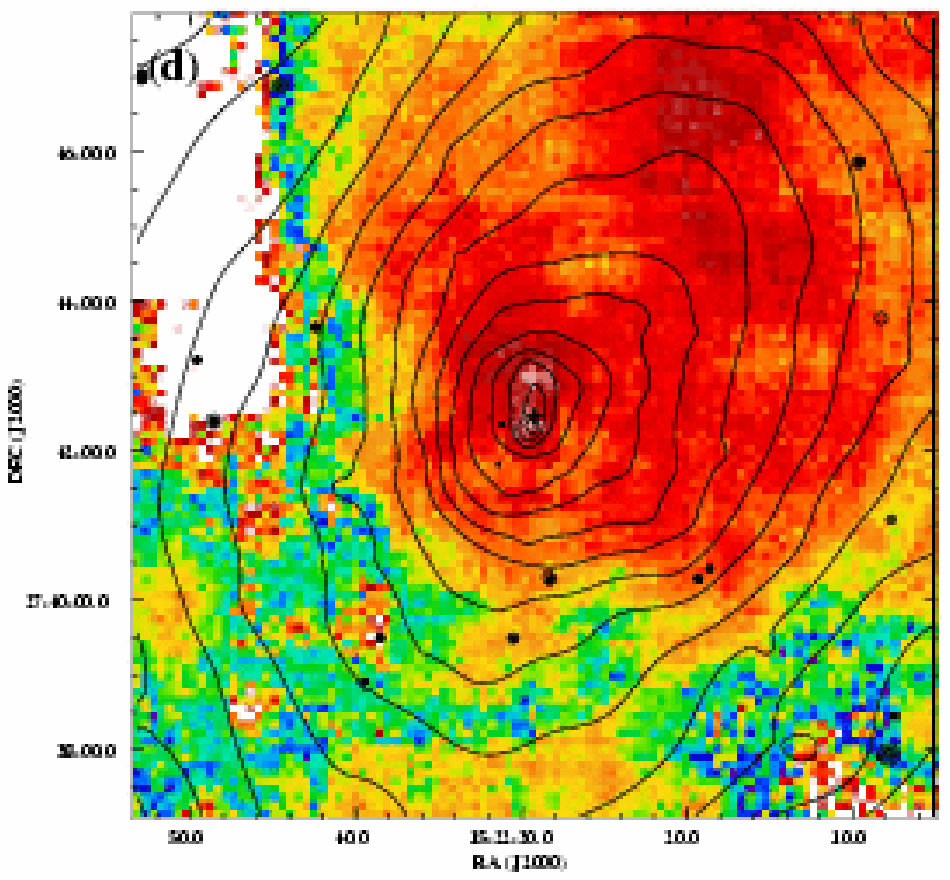}
\plottwo{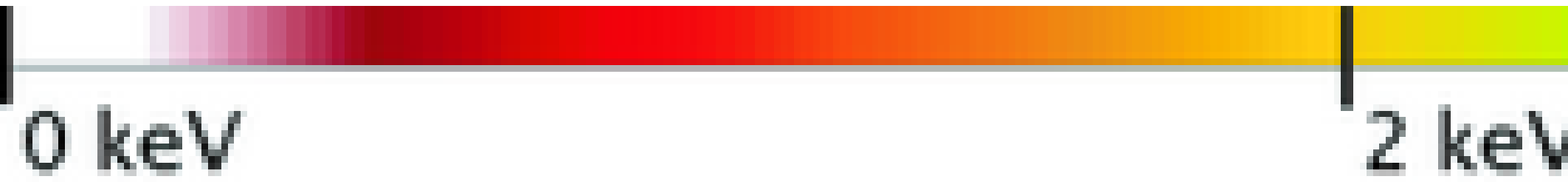}{f6f.eps}
\caption{Temperature maps derived from fitting the combined spectra
         from the two observations.
         The overlaid {\it Chandra} X-ray surface brightness contours
         are similar to those shown in Fig~\ref{fig:dss}.
         The cross denotes the position of the southern cD galaxy.
         {\it (a)} Map of a 10\farcm5$\times$10\farcm5
         field-of-view of the cluster.
         {\it (b)} Map of the central 4\farcm1$\times$4\farcm1
         of the cluster.
         The geometrical shapes illustrate the areas used for
         the extraction of cluster spectra ({\it solid lines})
         and for the local background to those spectra
         ({\it dashed lines}), as described in \S~\ref{sec:cool}.
         {\it (c-d)} Negative and positive, respectively, 90\%
         error maps of the temperature map shown in {\it a}.
         \label{fig:tmaps}}
\end{figure*}

Since the {\it Chandra} X-ray image of Abell 2065 shows a wealth of
structure and the previous {\it ASCA} temperature map showed evidence
for a shock structure to the south, we have constructed maps of the
projected temperature using the
Interactive Spectral Interpretation System,
ISIS\footnote{http://space.mit.edu/ASC/ISIS/}
\citep{ISIS}.
The temperature maps for two regions centered on the southern cD galaxy
have been computed and are presented in Figure~\ref{fig:tmaps}.
Fig.~\ref{fig:tmaps}{\it a} shows a region that stretches
10\farcm5~$\times$~10\farcm5 across, while the central
4\farcm1~$\times$~4\farcm1 of the cluster is presented
in Fig.~\ref{fig:tmaps}{\it b}.

The temperature calculations were performed as follows.
At each pixel position, the spectrum from each observation was extracted,
requiring that at least 800 net counts be included in the energy range of
0.7--6.5~keV.
The extraction region was a square box centered at the pixel center,
whose size varied having a maximum value of 500 ACIS-I pixels for
the larger scale map (Fig.~\ref{fig:tmaps}{\it a}) and $\sim$~300
ACIS-I pixels for the cluster center map (Fig.~\ref{fig:tmaps}{\it b}).
The box size was adaptively determined so that the number of observed
counts in the particular region, over the entire energy spectrum, was
between 1.5 and 2 times the nominal 800 counts, in order to account for
background subtraction as well as for the fact that the fitting occurs
over a limited energy range (0.7--6.5~keV).
For each observation, the background was extracted from the blank
sky background files, for the same energy band; for each pixel, the
region that was used was the same as the one used with the source
spectrum.
The spectra were combined in ISIS, requiring at least 25 counts
per bin, and subsequently fit with a thermal model with Galactic
absorption.
The abundance was held fixed at the value derived by fitting the
global spectrum with a thermal model (Table~\ref{tab:spec_global},
first row).

The temperature map of Figure~\ref{fig:tmaps}{\it a} reveals several
spatially separated regions of lower and higher projected temperatures.
The lowest temperatures are found to the NW and in a
limited region to the SE of the southern cD galaxy.
The entire region NW of the center of the cluster is moderately cool.
On the other hand, the hottest gas is at radii
of $\la$~3\arcmin\ SE of the southern cD.
The typical temperature to the NW of the cluster center is $\sim$~5~keV,
while the hotter SE region has much higher temperatures of $\sim$~10~keV.
These values are in good agreement with the values
\citet{MSV99} derived from {\it ASCA} data (regions
\#2 and \#4, respectively, in their paper).
These hot and cool regions seem to be separated by a
bow-shaped feature that runs from the NE to the SW of
the cD, at $\sim$~3\arcmin\ from it.
Especially to the SE of the cD, the temperature gradient
is quite steep, increasing by 7~keV over scales smaller
than an arcminute in some cases.
The temperature structure of that region and the high 
temperatures encountered in it, in combination with the
bow-shaped feature on the SE of the cD suggest the presence
of a shock front beyond $\sim$~3\arcmin\ to the SE of the
southern cD.

Figure~\ref{fig:tmaps}{\it c-d} show, respectively,
the negative and positive temperature error at each
pixel position, as calculated for 90\% error estimates.
The average error in these images is 1.56~keV and 1.70~keV,
respectively, for the negative and positive error images. 
It is obvious from these images, that the hot region to
the NE of the southern cD (Fig.~\ref{fig:tmaps}{\it a})
is associated with large errors, and thus it could be
an artifact of poor statistics.
Note that \citet{MSV99} did not find high temperatures
in this region (region \#5) with ASCA.

At smaller scales near the cluster center (Fig.~\ref{fig:tmaps}{\it b}),
the temperature map reveals at least two interesting structures.
The coolest gas in the map seems to be associated with the X-ray extension
to the south of the southern cD galaxy, and slightly displaced from
the center of this galaxy.
This would be unusual for a cool core in a relaxed system;
generally, they are very accurately coincident with the
center of the central cD.
As noted previously from Fig.~\ref{fig:radio}, both the thermal X-ray
and nonthermal radio plasma near the southern cD appear to be displaced
to the south.
The region just north of the southern cD seems to be warmer.
However, the tail further to the north is cooler
(the second coolest region in the cluster).

\subsection{Cool Cluster Regions} \label{sec:cool}

Whether the northern tail is actually a distinct gas concentration
from the southern extension is not clear from the error images
(Fig.~\ref{fig:tmaps}{\it c-d}).
The warmer material that bridges these two regions has a
projected temperature consistent with a monotonic radial
increase from the southern extension to the northern tail,
but also with a higher temperature than the surrounding
regions.

In order to address the significance of the apparent temperature
increase and to provide a more accurate assessment of the spectral
properties of these regions, we have extracted spectra and fit them
with a thermal APEC model with Galactic absorption.
The geometric regions defined for the extraction purposes
were such that they trace these features in the temperature
maps (Fig.~\ref{fig:tmaps}), the raw data image, as well as
the unsharp-masked image (Fig.~\ref{fig:radio}).
These regions are displayed in Fig.~\ref{fig:tmaps}{\it b}
by the solid-line geometrical figures.
In detail, the spectrum of the very cool southern extension was
extracted from a 12\arcsec\ radius circle centered 10\farcs1 south
of the southern cD center.
The spectrum of the warmer region between the two coolest cluster
regions (hereafter, the `bridge') was extracted from a rectangular
region, $\sim$~34\arcsec$\times$16\arcsec, elongated along the
east-west direction and centered 11\farcs1 to the north of the
southern cD.
Finally, the spectral extraction for the northern tail was performed
from a rectangular region centered 35\farcs7 to the north of the cD,
which was $\sim$~34\arcsec$\times$32\arcsec\ in size and run in the
east-west direction.

Since these regions are located at the center of the cluster, their
spectra are heavily contaminated by foreground and background X-ray 
emitting ICM along the line-of-sight.
Thus, accounting for the background has been accomplished in two different
ways: by extracting a local background from the events files themselves
and by using the blank sky background files.
The former case has the advantage that foreground/background cluster
emission is accounted for in a direct manner, and thus the derived
spectral properties should refer to represent the emission of the
small cool region near the cluster center.
The dashed-line rectangles in Fig.~\ref{fig:tmaps}{\it b}
illustrate the regions used for the local background.
In detail, the local background for the southern cool core was taken
from a rectangular region, $\sim 102\arcsec \times 49\arcsec$, elongated
in the east-west direction to avoid emission from the northern
structure and centered $\sim$~22\arcsec\ south of the southern cD,
excluding, of course, the circular region of the cool core
and point sources.
The local background for the bridge was extracted from regions
identical to and adjacently positioned to the east and west of
the rectangular region used with the spectrum extraction. 
For the tail, the background was extracted from 5 rectangular regions,
identical and contiguous to the one used for the spectrum extraction;
three of these regions were positioned to the north, and one each to the
east and west of the extraction rectangle.

On the other hand, when we use the blank sky background spectrum we
model the foreground and background emission by including a second
thermal component whose temperature and abundance were fixed to the
best-fit values from the global cluster X-ray spectrum.
The normalization of this component was allowed to vary.

The best-fit models for the southern cD extension
are shown in Table~\ref{tab:Treg}.
The two techniques yielded consistent results.
The blank-sky background yielded a better fit, although
with higher uncertainty in the best-fit values.
The abundances are essentially unconstrained with both
techniques.
To test if the use of the global cluster temperature is a good
approximation for the mean ICM temperature along the line-of-sight,
we fit the spectrum allowing the temperature and abundance of the
cluster component to vary.
The fit improved, but the line-of-sight thermal component is
now unconstrained, while the uncertainty of the cool core
temperature slightly increased.
The abundances are again unconstrained.

Consistent results are obtained for the bridge spectrum, as shown
in Table~\ref{tab:Treg}.
The local background technique yields a better fit and a higher
temperature.
Using the global spectrum temperature to model
the cluster emission did not improve the fit.
This suggests that the global spectrum may not provide an
adequate description of cluster emission along this line-of-sight.
Thus, we allowed the second thermal component to vary
but unfortunately the best-fit values for both thermal
components were unconstrained.
This, in combination with the uncertainty in the abundances in
both techniques, is a consequence of the low number of counts
in this region in both observations.

The two techniques yield consistent results for the northern tail,
as shown in Table~\ref{tab:Treg}.
The spectrum is adequately fit with either technique, though
the blank-sky background technique yielded a better fit.
Again, allowing the cluster component to vary slightly
improved the fit, but the computed best-fit parameters are more
uncertain.
As suggested by the temperature map (Fig.~\ref{fig:tmaps}),
the northern tail is cooler than the global temperature.
The metallicity is poorly constrained, but it appears to
be higher than the cluster average value.

Overall, these results suggest that the blank-sky background
technique provides slightly better fits, albeit with higher
uncertainties.
However, the temperature errors that either technique yields
are large enough to prevent a definitive assessment of the
significance of the temperature increase at the position of the
bridge.
In \S~\ref{sec:unequal} we argue that the northern tail is
a trail of gas that has been stripped from the southern cD
galaxy's cooling core.

\begin{deluxetable*}{llcccccc}
\tabletypesize{\rm\small}
\tablecaption{Spectral Fits for the Cluster's Coolest Regions
\label{tab:Treg}}
\tablehead
{
           &            
           &            
           &
 \multicolumn{2}{c}{Local Spectrum} &
           &
 \multicolumn{2}{c}{Cluster Emission}\\ \cline{4-5} \cline{7-8}
 Region     &
 Background &
 $\chi^2_r$ &
 $kT$       &
 Abundance  &
            &
 $kT$       &
 Abundance\\
           &
           &
           &
 (keV)     &
 (Solar Units) &
           &
 (keV)     &
 (Solar Units)
}
\startdata

 South cD Extension &
 Local &
 63.594/71 = 0.896 &
 $2.824^{+0.366}_{-0.322}$  &
 $1.096^{+0.716}_{-0.483}$  &
          &
 \dotfill &
 \dotfill  \\

 South cD Extension & 
 Blank Sky &
 58.765/70 = 0.839 &
 $2.073^{+0.893}_{-0.411}$  &
 $2.239^{+2.716}_{-1.016}$  &
       &
 5.519 &
 0.308  \\

 South cD Extension & 
 Blank Sky &
 58.281/68 = 0.857 &
 $2.094^{+0.961}_{-0.508}$  &
 $4.942_{-4.942}$  &
       &
 $5.148^{+9.137}_{-1.238}$  &
 $0.101^{+0.891}_{-0.101}$  \\

 Bridge to Tail & 
 Local &
 45.227/59 = 0.767 &
 $3.873^{+1.255}_{-1.010}$  &
 $1.591_{-1.242}$           &
       &
 \dotfill &
 \dotfill \\

 Bridge to Tail & 
 Blank Sky &
 53.381/58 = 0.920 &
 $2.128^{+2.313}_{-0.528}$  &
 $1.746_{-1.536}$  &
       &
 5.519 &
 0.308 \\

 Northern Tail  &
 Local &
 96.301/92 = 1.047 &
 $3.374^{+0.714}_{-0.466}$ &
 $0.916^{+1.293}_{-0.492}$ &
      &
 \dotfill &
 \dotfill \\

 Northern Tail &
 Blank Sky &
 90.393/91 = 0.993 &
 $2.324^{+1.591}_{-0.525}$  &
 $1.946_{-0.478}$  &
      &
 5.519 &
 0.308 \\

 Northern Tail  &
 Blank Sky &
 89.128/89 = 1.001 &
 $2.866^{+1.138}_{-0.758}$ &
 $0.864_{-0.410}$ &
      &
 $8.871_{-4.054}$ &
 $6.2_{-6.2} \times 10^{-6}$
\enddata
\end{deluxetable*}

\section{X-ray Surface Brightness Profile} \label{sec:sbprof}

The temperature map (Fig.~\ref{fig:tmaps}{\it a}) shows
a complex structure with an inner hot region and an outer
very hot region to the SE of the southern cD galaxy.
In addition, there are steep surface brightness gradients in the
same regions, displayed by the isophotes in Fig.~\ref{fig:dss};
these two results suggest the presence of a cold front and a
shock front in this region.

\begin{figure*}
\epsscale{0.42}
\rotatebox{90}{\plotone{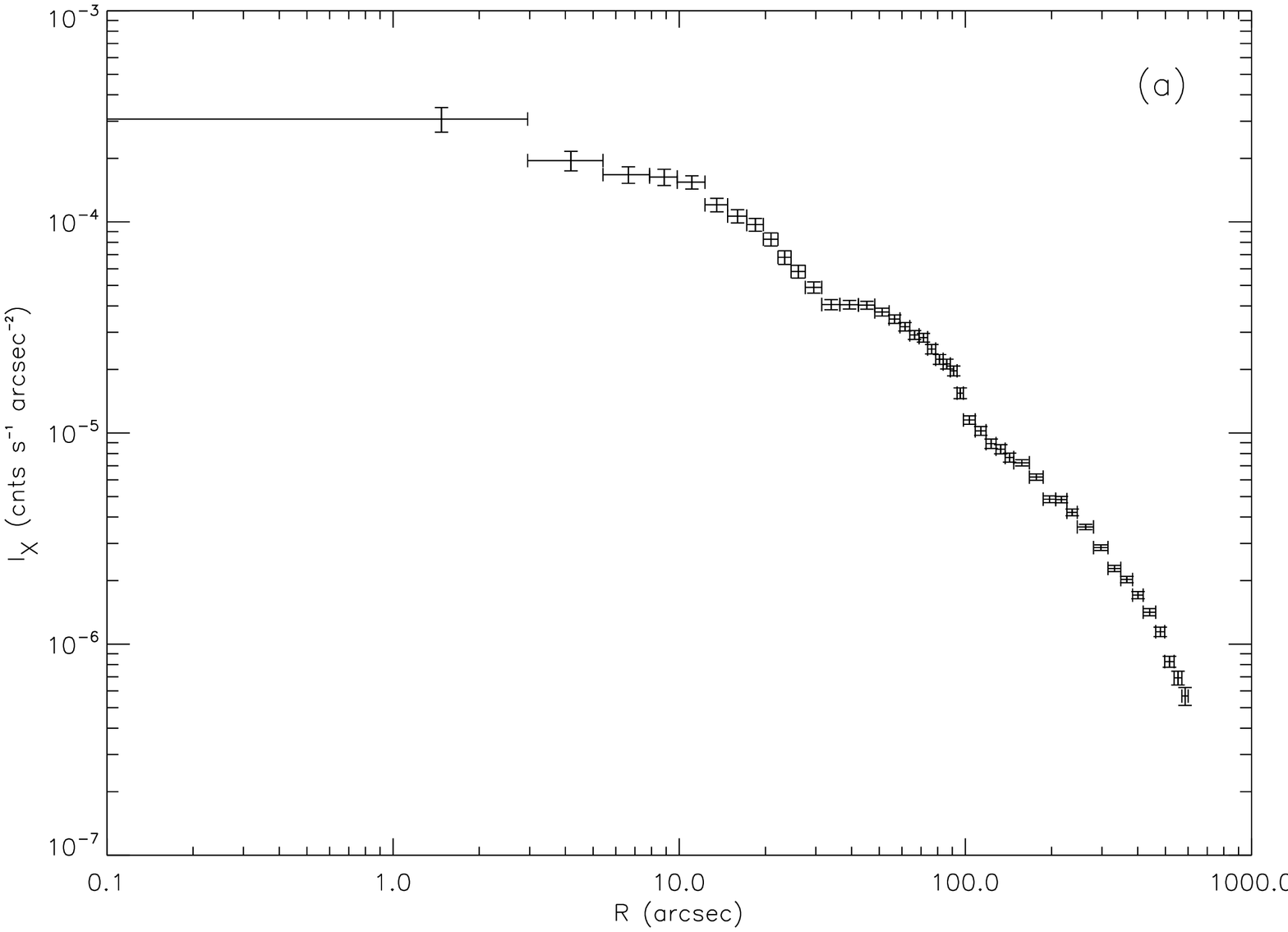}}
\rotatebox{90}{\plotone{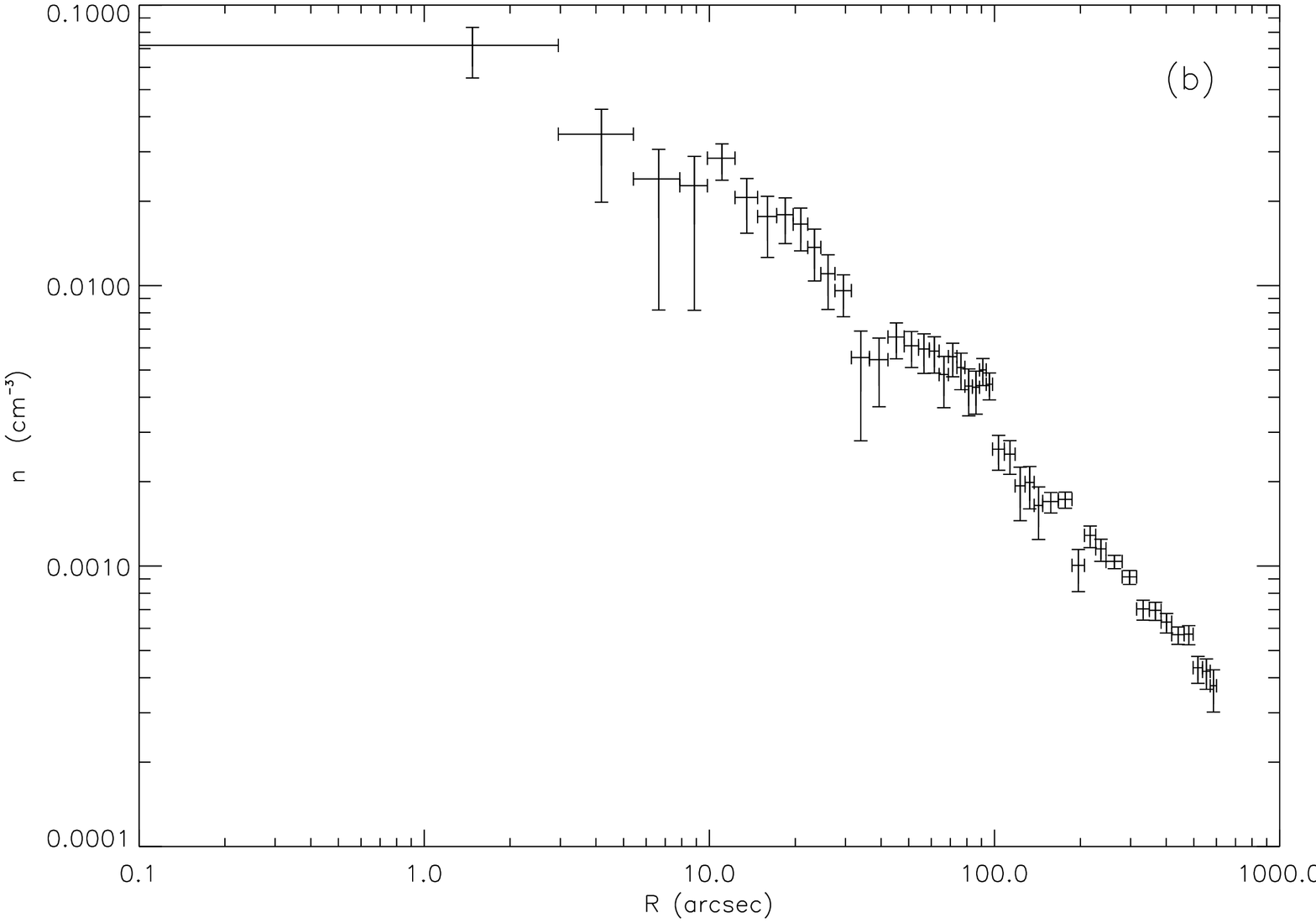}}
\caption{(a) Detailed surface brightness profile of the region
             SE of the southern cD galaxy.
             The rapidly declining inner portion of the plot
             (cluster's cool core) is succeeded by a possibly
             shocked region.
             The outer part of the profile is adequately fit with
             a $\beta$-model.
         (b) Local electron density distribution.
             Notice the clear presence of discontinuities at
             $\sim$30\arcsec\ and $\sim$100\arcsec.
             The discontinuity at $\sim$200\arcsec\ is artificial,
             as it reflects the moderate statistics obtained for
             that region, in which the chips gaps of the two
             observations overlap.
         \label{fig:sbprof}}
\end{figure*}

In order to better understand the gas physics in this cluster region,
we have extracted the surface brightness profile from a pie circular
aperture centered on the southern cD galaxy and lying to the SE.
This pie region stretches between 100\degr\ and 180\degr\ in PA
and out to $\sim$~10\arcmin\ from the center of the cD in radius;
a total of 45 annuli were used, their radial extent varying from
$\sim$~3\arcsec, for the innermost, to $\sim$~30\arcsec, for the
outermost.
The background was taken from the merged blank sky background file
(\S~\ref{sec:data}).
The resultant surface brightness distribution is presented in
Fig.~\ref{fig:sbprof}{\it a}.

The profile generally shows a rather smooth decline with radius.
Superposed on that, we find two interesting features.
First, there is a sharp decline over the inner part of the
profile, terminated by a discontinuity at $\sim$~30\arcsec\
from the center.
This discontinuity is associated with the cool southern cD extension.
This indicates that this feature is bounded by a cold front
at its southern edge.
Additionally, a region of excess X-ray emission, with respect to
the outer part of the brightness profile (see below), is found to
stretch from $\sim$~30\arcsec\ out to $\sim$~100\arcsec.
This is followed by a discontinuity, where the surface
brightness drops by a factor of $\sim$~1.7 over a radial
extent of 10\arcsec, reminiscent of surface brightness
jumps observed across shocks, as well as cold fronts.
Notice that this discontinuity is not associated with the
shock-like features seen in Fig.~\ref{fig:tmaps}{\it a}.
On the contrary, the outer 500\arcsec\ of the profile are
consistent with a $\beta$-model surface brightness distribution
($\beta = 0.63$ and $r_{core} = $ 218\arcsec\ or 292~kpc),
lacking evident shock signatures.

We have attempted to fit a cold front surface brightness model
to the inner part of the profile, but the excess X-ray emission
region that extends immediately outwards complicates the analysis.
Instead, in an attempt to better study the gas behavior across
the entire profile, as well as understand the nature of the
features seen in Fig.~\ref{fig:sbprof}{\it a}, we have deprojected
the surface brightness profile.
The assumptions that underlie this method are those of spherical
symmetry with respect to the gravitational potential peak, identified
in our case with the southern cD galaxy, and of uniform electron
density distribution within each shell.
The electron densities are presented in Figure~\ref{fig:sbprof}{\it b}.

The local density distribution clearly demonstrates the two features
seen in the projected profile of Fig.~\ref{fig:sbprof}{\it a}.
There is rather rapid central decline until $\sim$~30\arcsec,
where a discontinuity occurs.
The average density just inside the cold front region is
$\overline{n} = 2.67^{+0.59}_{-1.09} \times 10^{-2}$ cm$^{-3}$.
Then the density undergoes a shallow decline: it drops by a factor of 1.2
over a radial interval of $\sim$60\arcsec.
A dramatic density jump of $\sim$2 is observed at
$\sim$~100\arcsec, followed by a steeper decline.
Again, this jump is consistent with
being either a shock or a cold front.
The nature of this discontinuity is examined in \S~\ref{sec:shock}.
Finally, there seems to be another discontinuity at
$\sim$~3\arcmin\ from the southern cD, which coincides
with the position of the bow-shaped feature of
Fig.~\ref{fig:tmaps}{\it a}.
However, this jump also coincides with the region in our
merged raw data where the chip gaps of the two observations
overlap.
The magnitude of the density errors at the position of
this jump relative to the errors of adjacent values,
in combination with the absence of any strong
deviations from the smooth isothermal decline
in the surface brightness profile
(Fig.~\ref{fig:sbprof}{\it a}), indicate that the
jump is probably an artifact of low number
statistics at the position of the chip gaps.

\section{Spectral Deprojection} \label{sec:deproj}

Our interpretation of a shock propagating through the
cluster ICM has been stimulated by the temperature
map structure (Fig.~\ref{fig:tmaps}{\it a}).
However, the temperature map does not reflect the local
gas temperature (i.e. the gas at a specific distance from
the cluster center), rather the average temperature of the
X-ray emitting gas along the line-of-sight.
For instance, if a shock is indeed propagating beyond
$\sim$~3\arcmin\ SE of the southern cD, the projected
temperatures of the central cluster region, as shown in
the temperature map (Fig.~\ref{fig:tmaps}{\it b}), are
overestimates of the true local temperatures.

\begin{figure*}
\epsscale{0.65}
\plotone{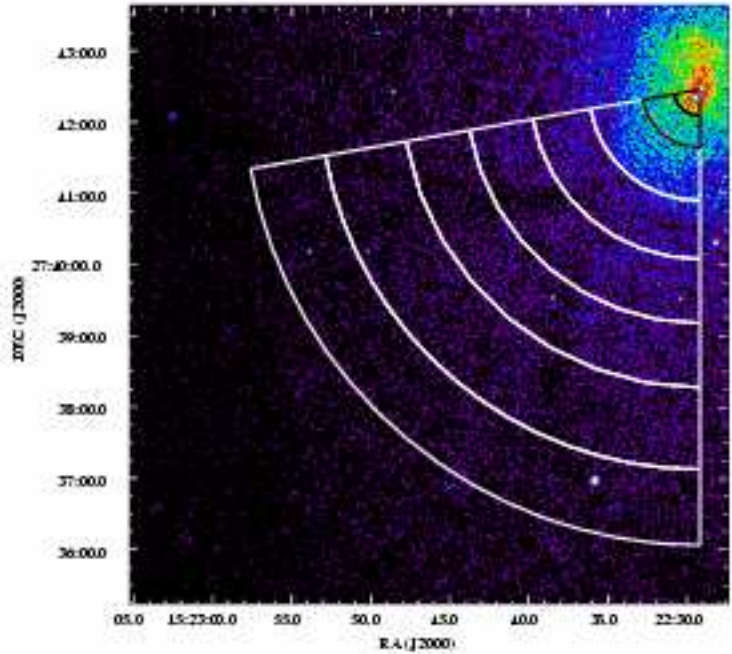}
\caption{Raw photon image in the range 0.7--6.5~keV
         of the $\sim$~8\farcm4$\times$8\farcm4 region
         SE of the southern cD.
         The image has been smoothed with a boxcar kernel
         $\sim$~1\arcsec\ and displayed in histogram
         equalization scale.
         The annuli show the regions used for the deprojection.
         The common vertex coincides with the center of the
         southern cD.
         The same overall region was used with the surface
         brightness profile deprojection.
         \label{fig:deprj:reg}}
\end{figure*}

In order to disentangle local from projected emission,
and gain deeper insight on the gas physics of this region,
we have deprojected the spectra of a set of regions.
Specifically, we have defined a set of 8 regions stretching
from the center of the optical cD out to a radius of
$\sim$~6\farcm4 and covering the same PA interval as
the regions used with the surface brightness deprojection
algorithm, namely from 100\degr\ to 180\degr.
These regions have been selected to fall on the ACIS-I detector
in both observations, and in particular to trace features of
the raw data image primarily, as well as of the temperature map.
For instance, the innermost region covers most of the area
used to examine the spectral properties of the southern
cD extension, discussed in \S~\ref{sec:cool}, while the
third annulus from the center traces the 100\arcsec\
brightness discontinuity.
They are displayed in Figure~\ref{fig:deprj:reg}.
In these regions, the total net counts from both
observations range between 1600 and 4200.

The spectra of these regions were extracted from each observation,
as described in \S~\ref{sec:spec}.
For each region, the flux weighted sum of the spectra from
the two observations for both the source and the background
were calculated, and a flux weighted average detector
response was produced.
The subsequent fitting was performed in the energy band
0.7--6.5~keV, grouping bins to contain at least 25 counts,
as needed.

The deprojection was carried out with XSPEC in two different
ways; in both, the shells' local emission was modeled with
a thermal model with Galactic absorption.
First, XSPEC's mixing model {\it projct} was used.
{\it projct} calculates the projection matrix of each of the
overlying shells on the area of each annulus and computes the
total projected emission on that annulus.
The fitting occurs simultaneously for all annuli and upon
completion the emission integral of the entire spherical
shell is returned, along with the best-fitting temperature
and abundance values.

Fitting all the spectra simultaneously has the effect that
the best-fit values of the outer annuli are affected by those
of the inner annuli.
To reduce the effects of this feature of the algorithm,
we have fit the individual spectrum of the outermost
annulus with an APEC thermal model with Galactic absorption.
Then, we used the obtained temperature and abundance values
to fix the spectral parameters of the outermost shell during
the deprojection and allowing all others to vary we obtained
a best-fit $\chi^2_r$ value of 0.96.
The abundance was rather poorly constrained.

An alternate approach was also followed.
In this, the spectral fitting occurred
successively for the different shells.
Starting with the outermost spectrum, the best-fit values
of the spectral parameters were determined.
Subsequently, the spectrum of the inner shell was fit
with two thermal components, one of which was fixed
to the best-fit values of the overlying shell, while
the relevant normalization was adjusted to account for
projection effects of the outer shell on the inner
region and for different exposures between the two
regions.
This was done for the spectra of all interior regions,
introducing thermal components to account for overlying
emission, as needed.
The abundance profile, again, was poorly constrained.

Comparing the two methods, it seems that, at least with our data,
the latter method yields temperature values with smaller errors.
The profiles presented in the following are based on these
results.

\subsection{Projected Temperature Profile} \label{sec:tproj}

\begin{figure}
\epsscale{0.9}
\rotatebox{90}{\plotone{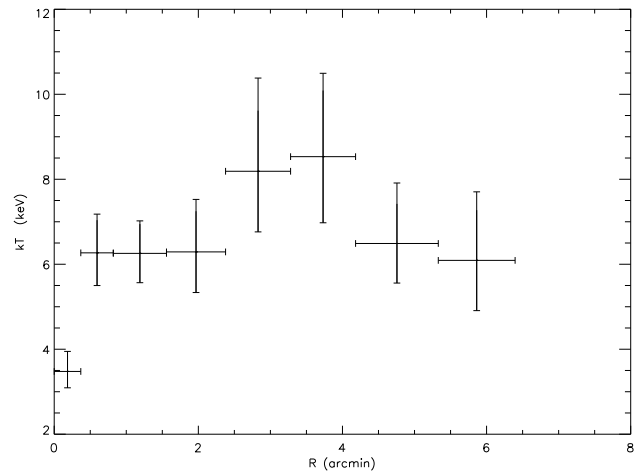}}
\caption{Projected temperature profile in the PA interval
         100\degr\--180\degr\ SE of the southern cD galaxy.}
         \label{fig:tproj}
\end{figure}

To check the consistency with the temperature map findings,
we have produced the projected temperature profile for these
regions, shown in Figure~\ref{fig:tproj}.
The cool gas at the center of the cluster is followed by
constant temperature gas out to a distance of $\sim$~2\farcm5
from the cD.
The temperature then rises by about 3~keV and maintains a
constant value within the errors for $\sim$~2\arcmin.
This rise, also seen in the temperature map, has motivated
the notion of shock propagation at that region.
Subsequently, it drops to values consistent with those
derived from {\it ASCA} data by \citet{MSV99}.
The temperature map does not show a similar trend at that
distance, but this is probably due to poor statistics at the
map edge.

Although the temperature rise and drop across 4\arcmin\
is reminiscent of shock heating, the shock hypothesis is
not supported by the detailed X-ray brightness profile
(Fig.~\ref{fig:sbprof}).
In the contrary, as was mentioned in \S~\ref{sec:sbprof},
the outer 500\arcsec\ are adequately fit with an isothermal
model, and bear no strong evidence for compression.
This hot, high entropy gas is then most likely a transient
merger-induced effect.
It apparently has a higher pressure than its surroundings
and it is expected to dissipate in a few hundred Myr
(a few sound crossing times).

\subsection{Deprojected Temperature Profile} \label{sec:tdeprj}

\begin{figure*}
\epsscale{0.42}
\rotatebox{90}{\plotone{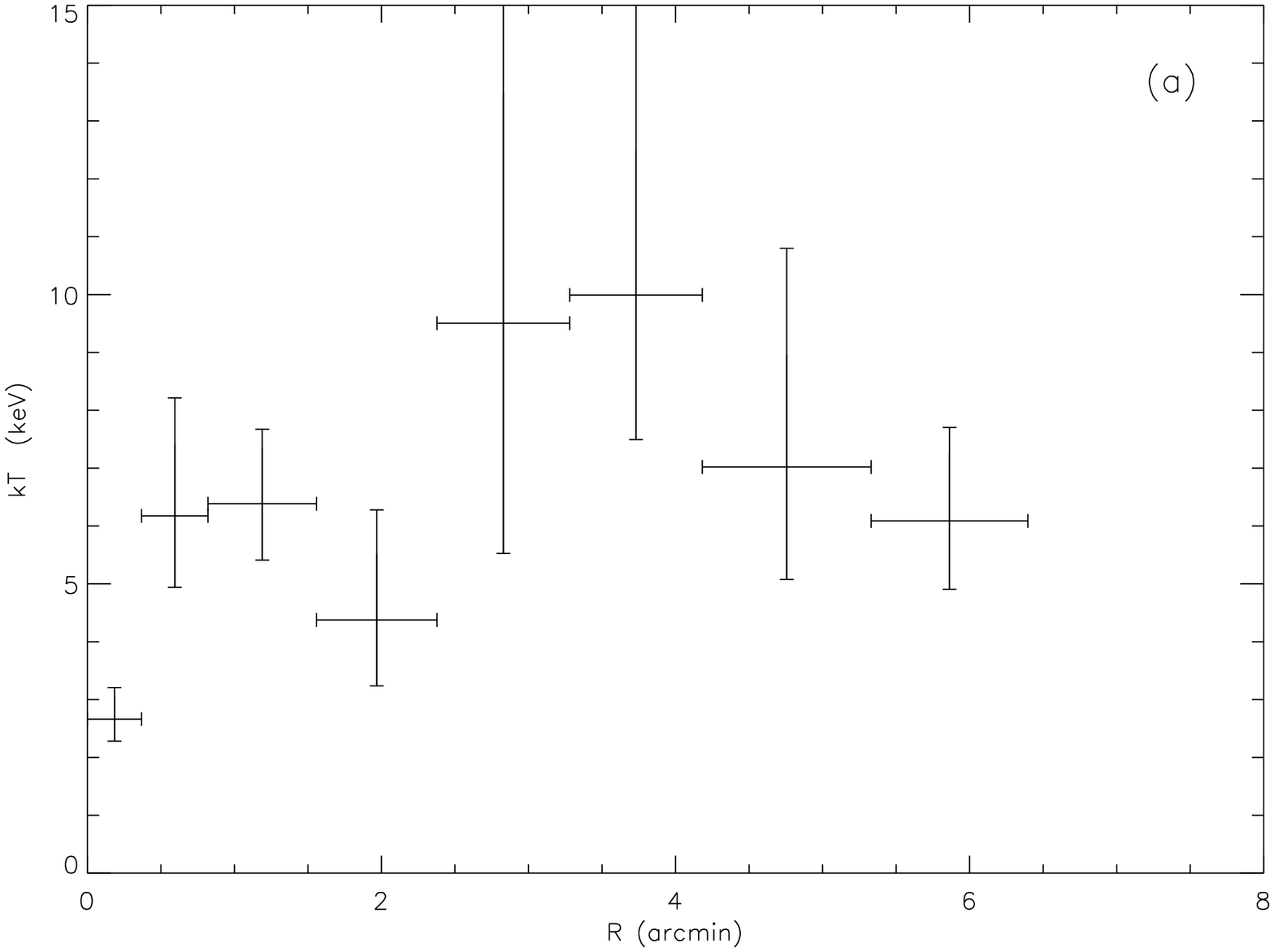}}
\rotatebox{90}{\plotone{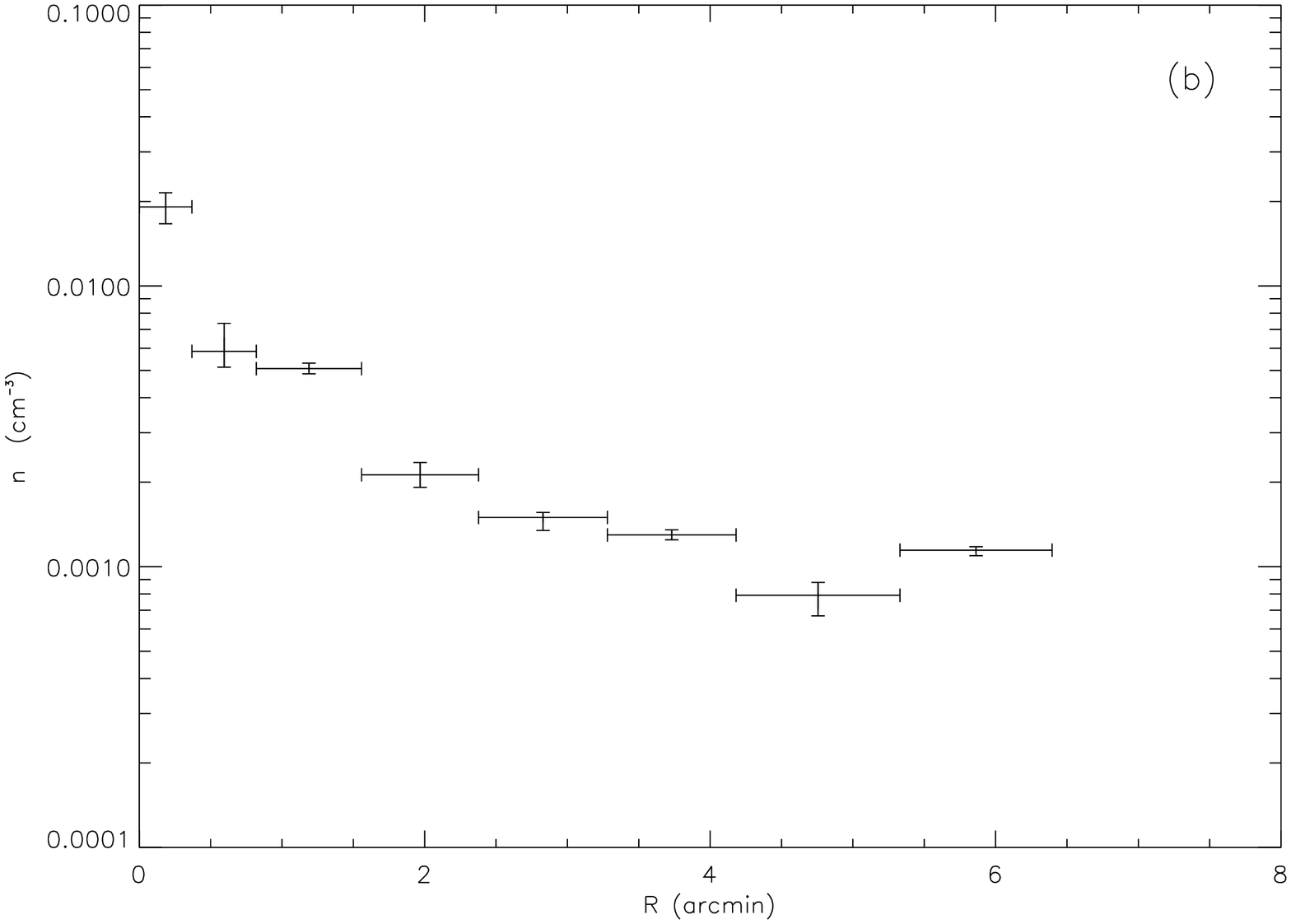}}
\rotatebox{90}{\plotone{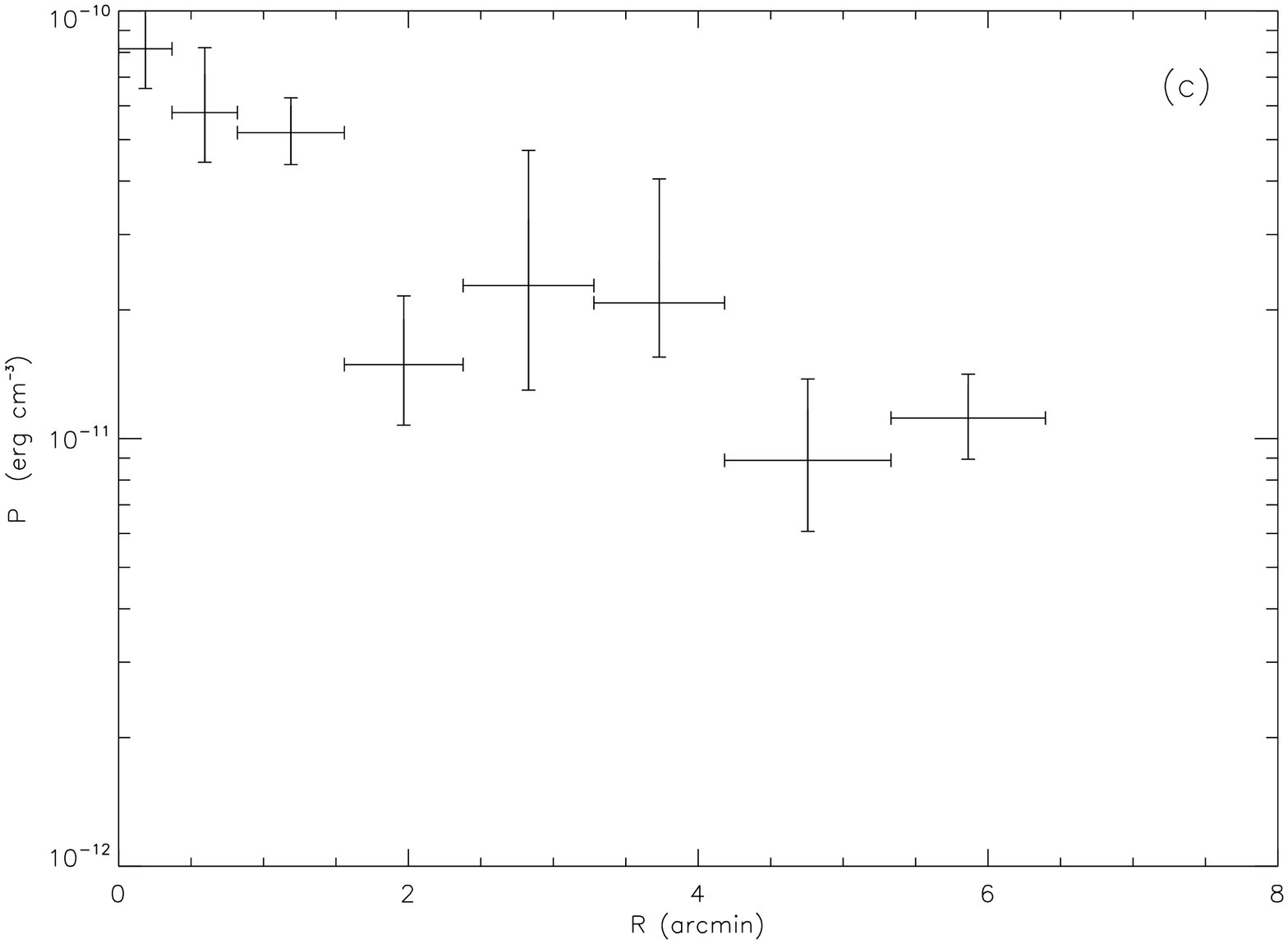}}
\rotatebox{90}{\plotone{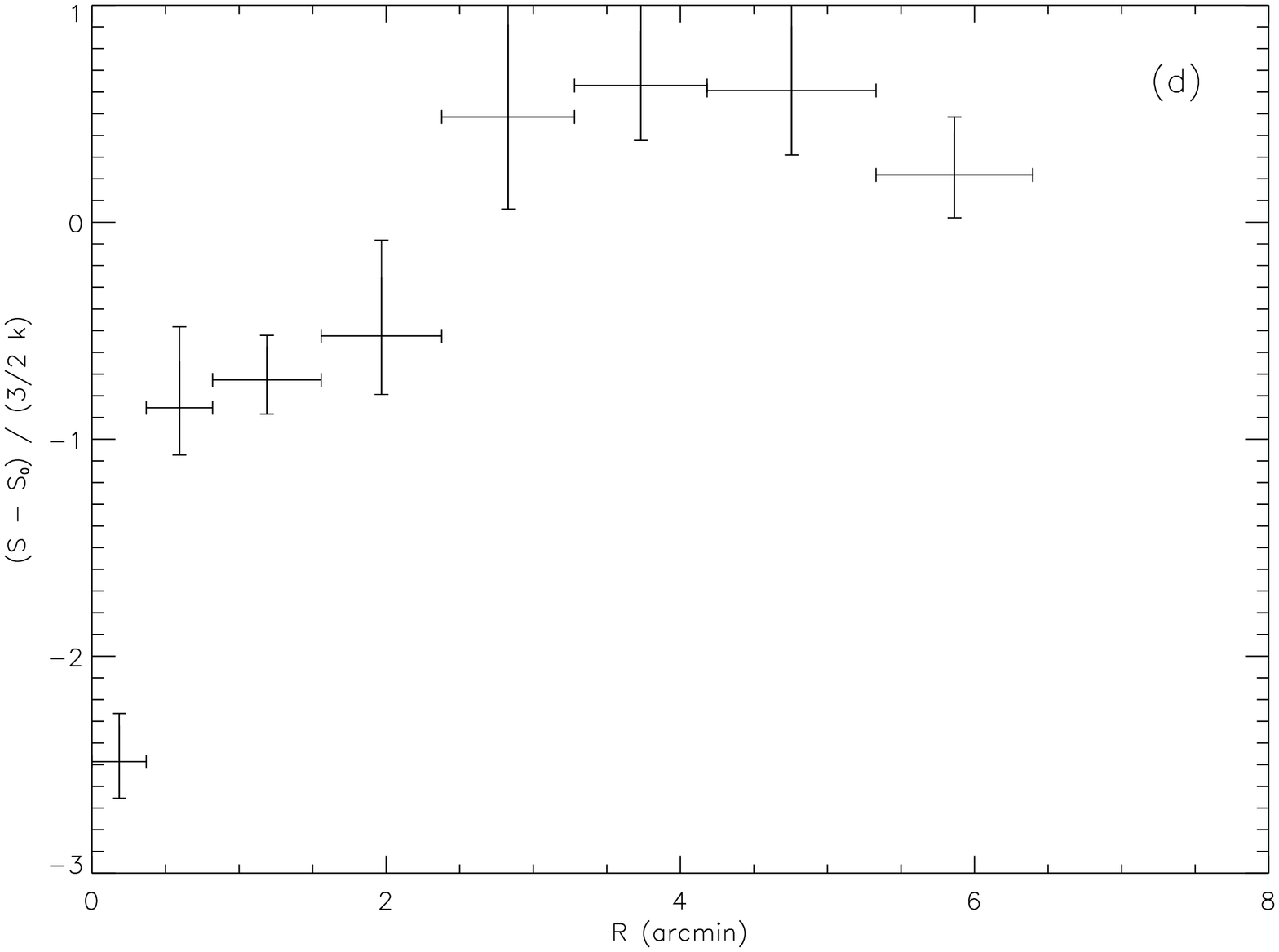}}
\caption{Deprojected ({\it a}) temperature, ({\it b}) density,
         ({\it c}) pressure and ({\it d}) specific entropy 
         profiles derived for the regions of Fig.~\ref{fig:deprj:reg}.
         Note that the quantity plotted in the entropy profile
         is the residual of the regions' entropy with respect
         to the average cluster entropy (see text for more details).
         \label{fig:deprj}}
\end{figure*}

The deprojected temperature profile is
presented in Figure~\ref{fig:deprj}{\it a}. 
The temperature in the central region is quite
low, as already indicated in Table~\ref{tab:Treg}.
The two high temperature points with very large
errors around 3\arcmin\ raise a concern.
If the temperature in these regions has been overestimated, then
deprojection might cause the temperatures of inner points to be
underestimated;
too much hot projected emission would have been subtracted from
these points.
In particular, the apparent temperature drop at about 2\arcmin\ might
then be an artifact of the overestimated temperature in the shells
just outside.
In any case, the drop at the position of the
brightness discontinuity (at $\sim$1\farcm6)
is not significant within the errors.

Given the highly uncertain temperature profile values,
we have decided to use the average value of 5.5~keV
to describe the gas temperature at the SE of the
brightness discontinuity. 
This value is consistent with temperature of the two outer
bins in the profile, and it is also consistent with the
the pre-shocked region temperature as estimated from {\it ASCA}
data \citep{MSV99}.
For error propagation purposes we adopt the values obtained
by \citet{MSV99} (region \# 6 of Fig. 3{\it c}, therein),
that is, $kT = 5.5^{+2.5}_{-1.5}$~keV.

\subsection{Density Profile}\label{sec:sbdeproj}

The definition for the thermal model normalization is that:
$K = \frac{10^{-14}}{4\pi D^2_A (1+z)^2}\int n_e n_p dV$,
where z the redshift to the cluster, $D_A$ is the angular
diameter distance at that redshift, $n_e$ and $n_p$ the
electron and proton number densities, respectively, and
the integration is performed over the shell's projected
volume.
Assuming uniform density distribution across each shell,
and that $n_e/n_p=1.21$, this equation may be inverted to
solve for the electron density.
The quality of our data was sufficient to allow
calculating accurate errors on the normalization
values.
We present the density profile in
Figure~\ref{fig:deprj}{\it b}.

The profile shows structure similar to the one revealed by the
higher resolution density distribution of
Fig.~\ref{fig:sbprof}{\it b}.
The density in the central 30\arcsec\ shell is high, and there is
an abrupt density drop of about a factor of $\sim$~4 outside of this
region.
The average density in the central 30\arcsec\ is consistent
with the average density derived from the detailed density
profile (Fig.~\ref{fig:sbprof}{\it b}).
Another discontinuity is seen at $\sim$~1\farcm6, coincident
with the position of the brightness discontinuity.
A monotonic decline beyond this point is interrupted by an
increase at the outmost shell, where the higher value is an
artifact of the assumption inherent to the deprojection
algorithm of no external emission.

\subsection{Pressure and Entropy Profiles}\label{sec:peprof}

From the temperature and density profiles and assuming
the ideal gas equation of state, we have calculated the
radial profiles of gas pressure and specific entropy.
They are presented in Fig.~\ref{fig:deprj}{\it c--d}, respectively.
The entropy was computed using the equation
$S = \frac{3 k}{2} \ln{\frac{T}{n^{2/3}}}$.
Note that the quantity plotted in the entropy profile is the residual
of the average cluster entropy from the regions' entropy,
$\Delta S = S - S_0$.
The global cluster entropy was calculated from the temperature
value of the best-fit model to the global spectrum, and the
average cluster density.

Note that the cool dense region at the center of the cluster
seems to be in pressure equilibrium with its surroundings,
while its entropy is the lowest in the cluster.
This suggests that the core is convectively stable.
The continuous pressure but greatly reduced entropy supports the
notion that the southern X-ray extension of the southern cD is
bounded by a contact discontinuity (a cold front).

The pressure shows a significant drop at $\sim$~1\farcm6, which
corresponds to the shock feature in Fig.~\ref{fig:sbprof}{\it a}.
There is no entropy decline between the same regions, as would be
expected for a shock, although the errors are large.

The outer 2\arcmin\ of the pressure profile are dominated by
large errors, but they overall seem to follow a declining trend.
Similarly, the outer portion of the entropy profile shows
rather elevated values, as expected from the projected
temperature profile (\S~\ref{sec:tproj}).

\section{Discussion} \label{sec:disc}

\subsection{Brightness Discontinuity at 100\arcsec} \label{sec:shock}

The detailed electron density profile (Fig.~\ref{fig:sbprof}{\it b})
shows a clear discontinuity at $\sim$~100\arcsec\ ($\sim$~140~kpc) to
the SE of the southern cD.
This discontinuity could be attributed to either a shock front
or a cold front at that position.
The critical parameter that would allow one to distinguish between
these two possibilities is the value of the temperature contrast
across the discontinuity.
The ratio of the temperature for material ahead of the front to
the temperature for material behind the front is expected to be
greater than unity for a cold front and less than unity for a
shock front.
The projected temperature profile bears no evidence for
any jump at that location, though the errors are significant. 
On the other hand, the deprojected temperature profile
(Fig.~\ref{fig:deprj}{\it a}) shows a jump consistent
with being greater than unity, but, as indicated earlier,
it may be an artifact of the deprojection.
Lacking strong evidence in either direction, we shall examine
the implications of both scenarios on the kinematics of the
merging system.

First, considering the shock possibility, the density
compression across the discontinuity
($\rho_2/\rho_1 = 1.918^{+0.297}_{-0.384}$)
can be used to constrain the kinematics of the merger. 
In principle, the Mach number ${\cal M}$ of a shock can be
determined by employing the Rankine-Hugoniot
shock equations (\citet{LL+59}, \S85) for density
or temperature jumps:
 \begin{equation}
 \frac{\rho_2}{\rho_1} = \frac{(1+\gamma) {\cal M}^2}{2+(\gamma-1) {\cal M}^2}
 ; \;
 \frac{T_2}{T_1} = \frac{(2\gamma {\cal M}^2 - \gamma+1){\cal M}}{\gamma+1}
 \frac{\rho_1}{\rho_2}
 \label{eqn:shock}
 \end{equation}
where the subscript 2 denotes material behind the shock (the
``post-shock'' region), while subscript 1 refers to material ahead
of the shock (the ``pre-shock'' region).
For intracluster gas, $\gamma = 5/3$.
Using the above compression value with the first part of
equation~(\ref{eqn:shock}), we confirm the presence of
a mildly supersonic outflow in this region
($\sim$~30\arcsec\--100\arcsec) with a Mach number of
${\cal M} = 1.66^{+0.24}_{-0.32}$ at the position of
the shock.
The observed temperature jump is much smaller than the
theoretical value, $T_2/T_1 =  2.78^{+0.88}_{-1.13}$,
although consistent within the errors.
Similarly, the observed and the theoretical values of the
pressure ratio are consistent within the errors, although
the pressure is not independent of either the density or
the temperature.

On the other hand, if the discontinuity is due to a cold front,
the velocity can be derived based on the stagnation condition
at the cold front \citep{VMM01a}, which gives the ratio of the
pressure at the stagnation point at the cold front ($P_{\rm st}$)
to that far upstream ($P_1$):
\begin{equation} \label{eq:pst}
\frac{P_{\rm st}}{P_1} = \left\{
\begin{array}{cl}
\left( 1 + \frac{\gamma - 1}{2} {\cal M}^2
\right)^{\frac{\gamma}{\gamma - 1}} \, , &
{\cal M} \le 1 \, , \\
{\cal M}^2 \,
\left( \frac{\gamma + 1}{2}
\right)^{\frac{\gamma + 1}{\gamma - 1}} \,
\left( \gamma - \frac{\gamma - 1}{2 {\cal M}^2} \,
\right)^{- \frac{1}{\gamma - 1}} \, , &
{\cal M} > 1 \, . \\
\end{array}
\right.
\end{equation}
Here, ${\cal M}$ is the Mach number of the cold front
relative to the upstream gas, and $\gamma = 5/3$.
We use the density at $\sim$~90\arcsec\ (Fig.~\ref{fig:sbprof})
and the temperature at 1\farcm2 (Fig.~\ref{fig:deprj}{\it a})
to approximate the pressure at the stagnation point.
We have chosen to approximate the pressure far upstream with
the density value at $\sim$~4\arcmin\ from the cD and the
average temperature 5.5~keV (\S~\ref{sec:tdeprj}).
Then, the pressure contrast is $\sim$~5, which implies that
the cold front is moving through the hotter gas with a Mach
number of ${\cal M} = 1.93^{+0.24}_{-0.19}$.
This should be viewed as an upper limit to the cold front
velocity, as applying equation~(\ref{eq:pst}) for regions
closer to the cold front suggests that the motion is still
supersonic, but with smaller Mach numbers.

It is important to realize that, regardless the ambiguity
of the nature of the discontinuity, these two scenarios
bear consistent kinematical implications.
In the following, we adopt the Mach number derived with
the shock condition for the motion of the material at
100\arcsec.

The sound velocity for a perfect gas, under adiabatic
conditions, is $c_s = \sqrt{\gamma k_B T/\mu m_p}$.
Taking the mean molecular weight of the ICM to be $\mu = 0.6$,
and using the average temperature 5.5~keV, the far upstream
sound speed is $\sim$~1200~km~s$^{-1}$.
Then, the velocity of the shock or cold front is
$v = {\cal M} c_s$, or $\sim$~2000~km~s$^{-1}$,
similar to the value derived by \citet{MSV99}.

\subsection{Stagnation Condition at the Cold Front} \label{sec:stagn}

The cool core surrounding the southern cD appears
to be bounded on its southern edge by a cold front.
An estimate of the merger relative velocity can also
be derived based on the stagnation condition at the
cold front, equation~(\ref{eq:pst}).
In this case, ${\cal M}$ is the Mach number of the cool core
relative to the upstream gas, and $\gamma = 5/3$.
As in the cold front scenario in \S~\ref{sec:shock},
the far upstream pressure has been approximated using
the density at a distance of $\sim$~4\arcmin\ SE of the
southern cD and the average temperature of 5.5~keV.
The stagnation point should be in pressure equilibrium
with the material inside the cool core; the density
immediately interior to the front ($r\approx$ 29\arcsec,
Fig.~\ref{fig:sbprof}{\it b}) and the deprojected core
temperature can be used to approximate the stagnation
pressure.
For this set of regions, the pressure ratio is about 4,
and the Mach number is ${\cal M} = 1.74^{+0.22}_{-0.18}$.
This value agrees with the Mach number of the discontinuity at 100\arcsec.
Especially in the case that the latter is indeed a cold front,
the discrepancy between the Mach number values of the two
structures could be understood as being due to acceleration
gained by the exterior front, as it propagates down a density
gradient.

\subsection{Survival of the Cool Core} \label{sec:coolcore}

The cool, low entropy, southern extension has been
identified with the cooling flow core of the southern cD.
The condition for survival of the cool core is (roughly) that
the ram pressure force of the incident ambient gas be less than
the gravitational restoring force of the potential of the cluster
core on the cooling core gas.
Assuming that the cooling core was initially in hydrostatic equilibrium,
the condition for the survival of the cool core is
\citep{FD91,MSV99,GLRB+02}:
\begin{equation}
\rho_{\rm sc} v_{\rm rel}^2 \la P_{\rm cc} -  P_{\rm am} \, ,
\label{eqn:ccsur}
\end{equation}
where $\rho_{\rm sc}$ is the cool core density of the infalling
cluster, $v_{\rm rel}$ is the relative velocity, $P_{\rm cc}$ is
the pressure at the center of the cool core, and
$P_{\rm am} \ll P_{\rm cc}$ is the ambient pressure of the main 
cluster around the cool core.
One concern is that this condition assumes that the cooling core gas is bound
to the dark matter potential of the main cluster;
in Abell~2065, the center of the cool core is displaced by
$\sim$~12\arcsec\ or 16~kpc from the center of the cD.
However, the fact that the cool core remains fairly intact probably implies
that this condition is not strongly violated.

The local density at 12\arcsec\ SE of the southern cD
(Fig.~\ref{fig:sbprof}{\it b}) has been employed for
the value of $\rho_{cc}$;
the central cooling core pressure $P_{\rm cc}$
was estimated utilizing the deprojected core
temperature.
The ambient pressure $P_{\rm am}$ has been estimated
from the density at 4\arcmin\ SE of the cD, and the
average temperature of 5.5~keV; it is negligible
compared to $P_{\rm cc}$ and is ignored in the following.
The choice for the infalling cluster initial central
density is more difficult to make.
Despite the ambiguity in the nature of the 100\arcsec\ 
discontinuity and of the high temperature region between
2\arcmin\ and 4\arcmin\, the density immediately outside
the 100\arcsec\ front has been taken as a good approximation
to $\rho_{\rm cc}$.
Indeed, even with this uncertainty, equation~(\ref{eqn:ccsur})
yields an upper limit to the relative subcluster velocity of
$v_{\rm rel} \lesssim $~1900~km~s$^{-1}$, in accordance with
the velocity derived from the kinematics of the 100\arcsec\
discontinuity.

\subsection{Dynamical Scenarios} \label{sec:dyn}

The results that have been presented above can be compiled into
a possible dynamical scenario concerning the history and the
stage of the merger.
In the following, we shall assume that the two cDs
are located at the centers of the two merging subclusters,
and that they are physically interacting with each other,
(that is, their close proximity is not a projection effect).
We will refer to the cluster centered on the southern cD
as the southern cluster, and to the one centered on the
northern cD as the northern cluster.

\subsubsection{Unequal Mass Merger} \label{sec:unequal}

We argue that the southern cluster is more massive than
its northern counterpart;
this is corroborated by the evidence presented above
and by numerical simulations, as it will be shown 
below in detail.
In this scenario, the northern cluster is falling
into the more massive southern cluster.
With the data, we can place an upper limit on the density contrast
of the two subcluster cores by assuming that the gas ahead of the
discontinuity at 100\arcsec\ serves as an undisturbed sample of
the initial northern cluster core conditions.
We have considered the density at the position of the X-ray
peak (at $\sim$~2\arcsec, Fig.~\ref{fig:sbprof}{\it b})
to be representative of the initial central density of the
southern cool core.
Then, the original density contrast has an upper limit of 27.

The absence of gas associated with the northern cD
suggests that either the northern cool core has been
disrupted during the merger or that the cluster did
not possess a cool core originally.
Although most subclusters have been observed to
possess cool cores, the latter possibility cannot
be dismissed.

The existence of the cold front at 30\arcsec\ and the
discontinuity at 100\arcsec\ to the SE of the merger
indicate that the southern cD is moving in this direction,
having already experienced core crossing.
Being the more massive and more dense part of the merger,
the southern cluster has swept away most of the northern
halo.
The shock that was triggered during the early merger stages has
displaced the northern cluster gas halo from the cluster's center
of mass and it is now projected to the SE of the merger, beyond
the discontinuity at 100\arcsec.

The temperature map structure (Fig.~\ref{fig:tmaps}{\it a}) is
similar to the structure revealed in the temperature map of the
1:16 unequal mass merger, with a density contrast of 50, in the
numerical simulations of \citet{GLRB+02}, when viewed 250~Myr
after core crossing.
This indicates that the density contrast may be higher
than the upper limit we determined above and that the
time since core crossing is probably a few hundred Myr.
This also suggests that the cluster has undergone
core crossing probably for the first time.

In this scenario, the cool core displacement in the direction of
motion can be understood in terms of the findings of \citet{Heinz03}.
As the northern cluster was falling into the southern cluster from the SE
(before core crossing), a shock wave was induced in the southern cluster.
The shock may have reached the dense, cool core of the southern
cluster, subjecting the latter to ram pressure stripping, which
removed material originally located at the outskirts of the core.
The generated pressure imbalance caused gas motions inside the
cool core that resulted in gas transportation against the shock
flow towards the contact discontinuity.

Finally, in the same scenario, the tail of cool gas that extends
from the southern cD to the north can be understood as the material
of the core outskirts that was removed by the incoming shock wave.
Apparently, this gas did not meet the survival criteria
set by equation~(\ref{eqn:ccsur}), and it has been stripped
from the cD.
The small temperature difference between the cool core and the
cold tail and their similar abundances (Table \ref{tab:Treg})
suggest that this may indeed be the case, since the shock could
displace as well as heat the gas to the observed values.

\subsubsection{An Alternate Possibility}

In \S~\ref{sec:images} the possibility was raised that the
southern cD galaxy may be moving to the NW, and that the observed
X-ray and radio extensions to the SE may be due to ram pressure
stripping.
In such a scenario, the two subclusters have not yet achieved
closest approach and it is possible that they may be merging
for the first time.
However, there are certain issues this possibility fails to
address.
The discontinuity to the SE of the southern cD galaxy could
be a manifestation of a shock wave, generated by the northern
cluster as it falls through the ICM of the southern cluster.
The induced ram pressure could be responsible for the
displacement of the southern cluster cooling flow from
the cD.
These considerations in combination with the fact that no
prominent shock front appears to the NW in our data, which
would be generated by the southern cluster as it free-falls
in the potential of its northern counterpart, suggest that
the northern cluster must be more massive than the southern.
Hence, it is remarkable that the cool core of the northern
subcluster does not manifest itself as a region of enhanced
X-ray emission in the X-ray brightness map of Fig.~\ref{fig:smooth}.
Additionally, it is much harder to explain the nature of the
cool tail that extends to the north of the southern cD, since
the numerical experiments of \citet{Heinz03} suggest that a shock
front driven into a cooling flow would displace the latter
in the incoming shock's direction, but it would not distort
its geometry to form an elongated leading structure.
The possibility that this gas may have been expelled from
the northern cD is not favored by the temperature consistency
between the gas along the tail and the gas of the southern
extension.

\section{Summary} \label{sec:sum}

We have presented the results of a {\it Chandra} observation
of the merging cluster of galaxies Abell 2065, which include the
temperature maps of the central 10~arcmin$^2$ of the cluster
emission and the spectral properties of the cooling flow of
the southern cluster as well as a plume of cold gas emerging
from the cD to the north.
Evidence of shocks appear in the temperature maps to the SE
of the merger and the deprojected density distribution of that
region indicates the presence of a supersonic flow to the SE,
with ${\cal M} \approx$~1.7.
However, the temperature data are not of sufficient accuracy to
allow distinguishing between the shock wave and the cold
front interpretation of this discontinuity.
The cooling flow is displaced to the SE of the cD. 
At its southern edge, the cooling core appears to be bounded by
a cold front.

Putting all these together, we propose that Abell 2065
is an unequal mass merger.
The northern cluster seems to have fallen in the more
massive southern cluster from the SE probably for the 
first time and having lost its gas content it is now
moving to the NW, where it is now seen.

\acknowledgments

We would like to thank the anonymous referee for helpful remarks.
M.C. would like to thank Maxim Markevitch, Adrienne Juett and
Gregory Sivakoff for helpful comments and discussions.
Support for this work was provided by the National Aeronautics and
Space Administration primarily through {\it Chandra} award GO2-3159X,
but also through GO2-3160X, GO3-4160X, GO4-5133X, GO4-5137X, and GO5-6126X,
issued by the Chandra X-ray Observatory, which is operated by the
Smithsonian Astrophysical Observatory for and on behalf of NASA
under contract NAS8-39073.


\begin{thebibliography}{}


\bibitem[Abell et al.(1989)]{ACO89}
Abell, G. O., Corwin, H. G., \& Olowin, R. P.\
1989, ApJS, 70, 1

\bibitem[Abramopoulos \& Ku(1983)]{AK83}
Abramopoulos, F., \& Ku, W.~H.-M.\
1983, \apj, 271, 446

\bibitem[Arnaud(1996)]{XSPEC}
Arnaud, K. A.
1996, in ASP Conference Proceedings,
Astronomical Data Analysis Software and Systems V,
ed.\ Jacoby G. and Barnes J.,
(San Francisco: ASP), 17

\bibitem[David et al.(1994)]{DJFD04}
David, L.~P., Jones, C., Forman, W., \& Daines, S.\
1994, \apj, 428, 544 
 
\bibitem[David et al.(1993)]{DSJ+93}
David, L.~P., Slyz, A., Jones, C., Forman, W., Vrtilek, S. D., \& Arnaud, K. A.\
1993, ApJ, 412, 479

\bibitem[Fabian \& Daines(1991)]{FD91}
Fabian, A. C., \& Daines, S. J.\
1991, MNRAS, 252, 17P

\bibitem[Fabian et al.(2003)]{2003MNRAS.344L..43F}
Fabian, A.~C., Sanders, J.~S., Allen, S.~W., Crawford, C.~S.,
Iwasawa, K., Johnstone, R.~M., Schmidt, R.~W., \& Taylor, G.~B.\
 2003, \mnras, 344, L43 
 
\bibitem[Fujita et al.(2004)]{FSW04}
Fujita, Y., Suzuki, T. K., \& Wada, K.\
2004, ApJ, 600, 650

\bibitem[G\'{o}mez et al.(2002)]{GLRB+02}
G\'{o}mez, P. L., Loken, C., Roettiger, K., \& Burns, O.\
2002, ApJ, 569, 122

\bibitem[Heinz et al.(2003)]{Heinz03}
Heinz, S., Churazov, E., Forman, W., Jones, C., \& Briel, U. G.\
2003, MNRAS, 346, 13

\bibitem[Houck et al.(2000)]{ISIS}
Houck, J. C., Denicola, L. A.
2000, in ASP Conference Proceedings,
Astronomical Data Analysis Software and Systems IX,
ed.\ Nadine Manset, Christian Veillet, \& Dennis Crabtree,
(San Francisco: ASP), 591

\bibitem[Ikebe et al.(2002)]{IPB+02}
Ikebe, Y., Reiprich, T. H., B\"ohringer, H., Tanaka, Y., \& Kitayama, T.\
2002, A\&A, 383, 773

\bibitem[Landau \& Lifshitz (1959)]{LL+59}
Landau, L., D., \& Lifshitz, E. M.
1959, Fluid Dynamics, (Reading, Massachusetts: Addison-Wesley)

\bibitem[Liedahl et al.(1995)]{mekal}
Liedahl, D. A., Osterheld, A. L., \& Goldstein, W. H.\
1995, ApJL, 438, 115

\bibitem[Markevitch et al.(1998)]{MFSV+98}
Markevitch, M., Forman, W. R., Sarazin, C. L., \& Vikhlinin, A.\
1998, ApJ, 503, 77

\bibitem[Markevitch et al.(1999)]{MSV99}
Markevitch, M., Sarazin, C. L., \& Vikhlinin, A.\
1999, ApJ, 521, 526

\bibitem[Peres et al.(1998)]{Peres}
Peres, C. B., Fabian, A. C., Edge, A. C., Allen, S. W., Johnstone, R. M.,
 \& White, D. A.\
1998, MNRAS, 298, 416

\bibitem[Postman et al.(1988)]{PGH}
Postman M., Geller, M. J. \& Huchra, J. P.\
1988, AJ, 95 267

\bibitem[Ricker \& Sarazin(2001)]{RS+01}
Ricker, P. M., \& Sarazin, C. L.\
2001, ApJ, 561, 621

\bibitem[Smith et al.(2001)]{apec}
Smith, R. K., Brickhouse, N. S., Liedahl, D. A., \& Raymond, J. C.\
2001, ApJ, 556, 91

\bibitem[Vikhlinin et al.(2001)]{VMM01a}
Vikhlinin, A., Markevitch, M., \& Murray, S. S.\
2001, ApJ, 551, 160

\bibitem[White(2000)]{Whi00}
White, D. A.\
2000, MNRAS, 312, 663

\end{thebibliography}
\end{document}